\documentclass[%
pre,
superscriptaddress, 
preprint,
preprintnumbers,
 amsmath,amssymb,
 aps,
15pt,
]{revtex4-2}

\usepackage{graphicx}
\usepackage{dcolumn}
\usepackage{bm}
\usepackage{todonotes}
\usepackage{caption}

\usepackage{newtxtext}
\usepackage{newtxmath}
\usepackage{natbib}
\usepackage{multirow}
\usepackage[utf8]{inputenc}
\usepackage{float}
\usepackage{soul}
\usepackage{hyperref}

\hypersetup{
    colorlinks,
    linkcolor={blue},
    citecolor={blue},
    urlcolor={blue}
}

\begin{document}

\title{Large-scale patterns of small-scale vorticity interactions foster moist convection during cyclogenesis}

\author{Shruti Tandon}
\affiliation{Centre of Excellence for studying Critical Transitions in Complex Systems, Indian Institute of Technology Madras, Chennai 600036, India}
\affiliation{%
Department of Aerospace Engineering, Indian Institute of Technology Madras, Chennai 600036, India
}
\author{Apoorva Singh}
\affiliation{Centre of Excellence for studying Critical Transitions in Complex Systems, Indian Institute of Technology Madras, Chennai 600036, India}
\affiliation{%
currently at: Department of Applied Physics and Science Education, Eindhoven University of Technology, Netherlands
}
\author{B. N. Goswami}
\affiliation{ST Radar Centre, Gauhati University, Guwahati-781014, India}
\author{R. I. Sujith}
\email{sujith@iitm.ac.in}
\affiliation{Centre of Excellence for studying Critical Transitions in Complex Systems, Indian Institute of Technology Madras, Chennai 600036, India}
\affiliation{%
Department of Aerospace Engineering, Indian Institute of Technology Madras, Chennai 600036, India
}

\maketitle
Understanding the self-organization of small-scale moist convection at synoptic scales during tropical cyclogenesis is essential. We examine the role of vorticity interactions in dynamical organization of moist convection during cyclogenesis in the Bay of Bengal. Using time-varying complex networks, we encode local vorticity correlations between nodes (locations), only within a spatial proximity of sub-meso length scales. Links are established using short-time correlation of vorticity fluctuations. Spatial patches of high network connectivity depict regions of locally coherent vorticity dynamics. We find that various patches of high network connectivity arrange into extended structures forming identifiable large-scale patterns. Examining the co-evolution of moist thermodynamics parameters with local vorticity coherence, we discover a unique spatial correlation between the convective processes occurring at far away locations, owing to self-organization. We find that such spatial correlation are evident only for depressions that form cyclones, and not otherwise. We discover the physical mechanism of self-organization using modal analysis of network connectivity and vorticity budgets. In regions identified as most influential by the modal analysis, we find that vertical vorticity is enhanced locally as well as advected horizontally outwards to other locations. Such horizontal advection of vertical vorticity induces nonlocal interactions and couples small-scale convective processes across a large domain. We show that, vertical vorticity transported from regions of high convective potential help sustain strong updrafts and weaken convective inhibition in other regions. Thus, spatial interdependence is introduced between small-scale convective processes at different locations, thus explaining the self-organization of moist convection during cyclogenesis.

\section{Introduction}

Predicting the likelihood of the formation of a tropical cyclone in an environment of high moisture and vorticity is not trivial. Planetary scale tropical perturbations initiate the formation of low-pressure centers frequently. However, only a few of these low-pressure centers develop into identifiable depressions. Even fewer of these depressions intensify into cyclones. Tropical cyclones occur frequently in tropical ocean basins \citep{klotzbach2006trends,wu2022cycloneintensity} and the occurrence of one cyclone can cause close to a hundred million dollars of economic losses on an average \citep{wmo_tropical_cyclone, kunze2021unraveling, krichene2023social}. A cyclone attains its peak intensity depending on the environment feeding its sustenance \citep{gray1998formationrev, emanuel2003tropicalrev, montgomery2017recent}. Some cyclones exhibit rapid intensification and cause much more damage than cyclones that reach similar strengths gradually. For example, Amphan (2020), a category 5 cyclone \citep{imd2020amphanpr,bhowmick2020cycloneamphan}, caused much higher precipitation inducing devastating floods, and led to one of the largest economic losses (more than $10$ billion US dollars) compared to most cyclones that occurred in the North Indian ocean \citep{imd2020amphanpr,wmo2021climate, biswas2022tropical,chatterjee2021complete}. 

Today, we understand the structure of a mature cyclone \citep{emanuel2003tropicalrev, montgomery2017recent}, and can predict the path of cyclones fairly well \citep{chan2005physicsrev}. However, distinguishing tropical depressions that develop into a cyclone from those that do not, is still an open challenge. This is because several physical processes that aid cyclogenesis and the relative contribution of each of these processes are not well established. 
In order to identify when a depression will (or not) form a cyclone, we need to improve our understanding of how vorticity interactions aid the dynamical organization of cumulus convection during cyclogenesis. Here, we use the lens of complex systems theory \citep{kauffman1995home, ladyman2013complex, ottino2003complex} and view the formation of a cyclone as the self-organization of local thermo-fluid interactions, leading to the emergence of large-scale coherent patterns amidst a turbulent atmosphere. Using this perspective, we find that depressions form cyclones when large-scale patterns that emerge from local vorticity interactions organize the moisture convection and convergence across a large-scale region.

\begin{figure}
    \centering
    \includegraphics[width=1\linewidth]{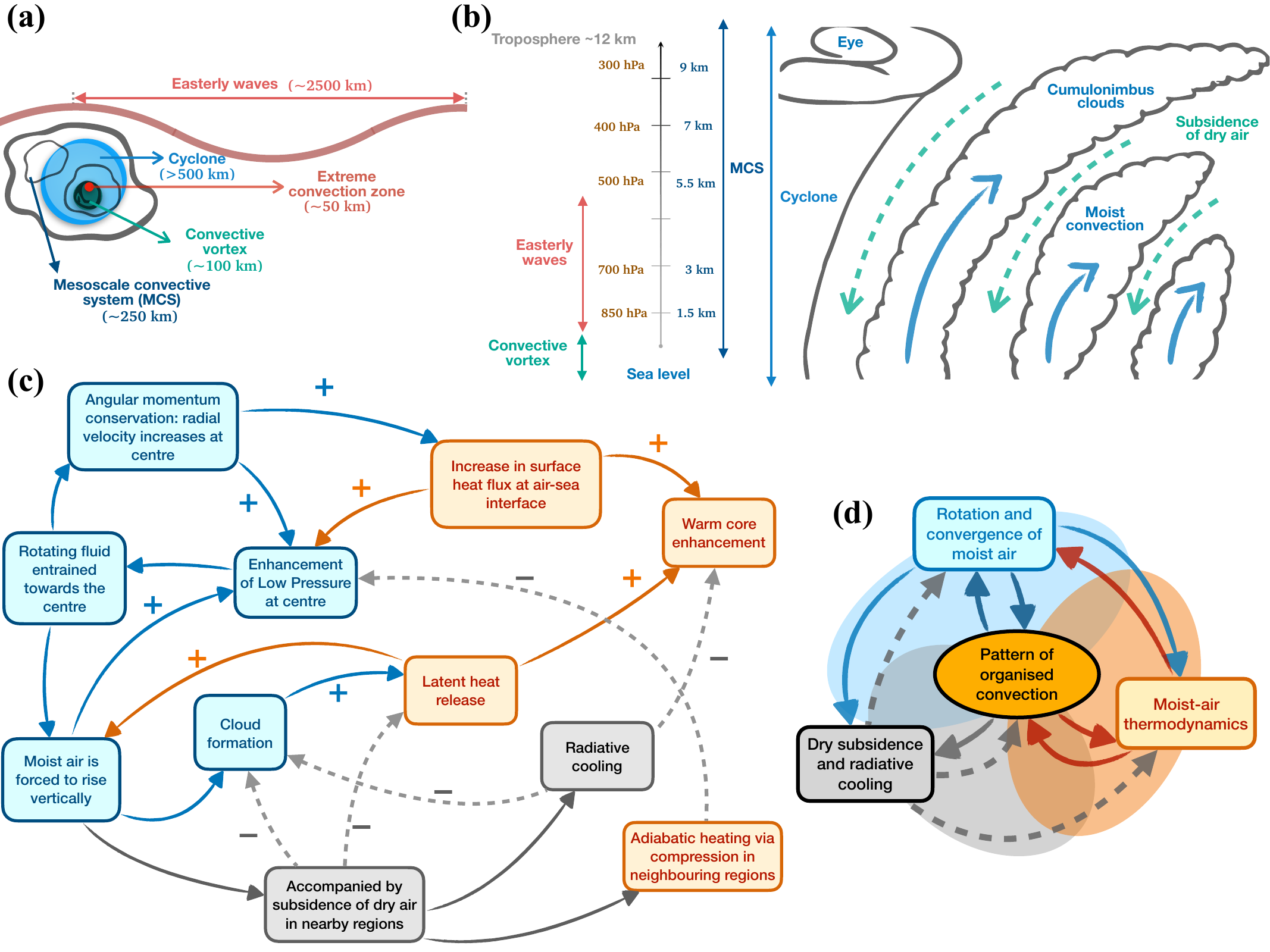}
    \caption{Diagram representing (a) the horizontal and (b) the vertical spatial extents (not to scale) of tropical cyclones as compared to easterly waves, mesoscale convective systems (MCS), and low-level convective vortices associated with lower tropospheric cloud-generated vorticity. Extreme convection zones can form within an MCS. A tropical cyclone is $> 500$ km in diameter and spans from sea surface to troposphere ($\sim 12$ km in height). (c) A schematic flow chart showing the various thermo-fluid feedback interactions that sustain and foster genesis and intensification of tropical cyclones. (d) A diagrammatic representation of interdependence between the pattern of organized convection, and thermo-fluid feedback interactions including radiative cooling in regions of dry subsidence, moist air thermodynamics and convergence of rotating moist air. A tropical cyclone is essentially a self-sustained and intensified pattern of organized convection.}
    \label{fig_schema}
\end{figure}

Multiple theories of cyclogenesis have been proposed over the past several decades involving various thermo-fluid processes and feedback (figure \ref{fig_schema}(c)) that occur at various length scales (figure \ref{fig_schema}(a,b)).
The earliest explanation came from imagining a cooperative interaction between multiple regions of moist air convergence and a large cyclonic vortex, referred as the theory of conditional instability of the second kind (CISK) \citep{charney1964growth,ooyama1969numerical}. According to CISK, a large synoptic scale vortex organizes the wind and drives large-scale wind convergence in the lower troposphere, which then induces convection of warm moist air. Such convection leads to cloud formation releasing latent heat energy thus providing feedback to intensify the vortex. Thus, CISK implies that accumulated convective available potential energy (CAPE) is released rapidly due to convection induced by convergence in the lower troposphere during cyclogenesis. CAPE is a thermodynamic parameter that is a measure of energy available to lift a moist parcel vertically and hence a measure of potential for convective instability in the atmosphere. However, high CAPE alone does not imply convection; high CAPE must overcome the local convection inhibition (CI) energy associated with downdrafts and radiative feedback (as shown in figure \ref{fig_schema}(c)). The selective release of built-up CAPE against CI does not explain self-sustained convergence and large-scale organization in the flow. 

Subsequently, \citet{arakawa1974interaction} argued that CAPE generally attains a quasi-equilibrium in tropical atmosphere thus maintaining a balanced gradient on a large-scale. Thus, the sudden rapid consumption of CAPE due to convection induced by large-scale convergence is an inadequate explanation. To overcome this deficit, another theory proposed by \citet{emanuel1986air} explained that cyclonic wind generated by flow instability in the atmosphere enhances the heat transfer from sea to air. Such wind induced surface heat energy (WISHE) leads to massive evaporation and adds moisture and energy to the cyclonic system. The theory implies that even though CAPE is consumed rapidly during the initial stage through rapid convection, cyclogenesis is sustained through WISHE and moisture fluxes alone. 
However, CAPE can increase or remain significant during cyclone intensification \citep{miyamoto2013transition,lee2018role} and formation \citep{schecter2011evaluation}. Further, not all tropical cyclone observations and simulations corroborate the description of vanishing CAPE before cyclogenesis \citep{lee2018role,montgomery2009tropical}. Moreover, randomness in the initial condition of moist convection can lead to drastic changes in the large-scale organization and the probability of cyclogenesis \citep{zhang2009effects}. 
\citet{lee2018role} have shown that such reduction of CAPE in the WISHE theory leads to inhibited intensification, and that an increase in CAPE in the radially outward direction is favorable for greater intensification of the storm. Despite the success of these theories in explaining some of the thermo-fluid interactions that occur during cyclogenesis, the role of CAPE in the early stages of cyclone formation remains elusive.

The works of \citet{enagonio2001tropical} and \citet{montgomery2006vortical} have implied that incorporating the role of CAPE is essential to explain cyclogenesis in contrast to the WISHE theory. They demonstrated that the intensification of an incipient vortex into a cyclone involves the formation of deep rotating convective structures, namely vortical hot towers (VHTs). Strong updrafts occur in small regions ($\approx 50$ km) of near saturation conditions forming clouds (extreme convection zones demarcated in figure \ref{fig_schema}(a)), as also explained earlier by \citet{gray1998formationrev}. Such vertical surge high humidity wind is triggered in regions of concentrated vorticity repeatedly prior to cyclogenesis. Rising moist air forms clouds at the cumulus scale (of the order~$\approx 1$ km), releasing latent heat and energizing the moist air to rise further in altitude. Due to deep convection, VHTs are associated with regions of high CAPE and subsequently consume CAPE locally generating intense vorticity anomalies and intensifying the vorticity at relatively smaller (sub-meso) scales. As a result, deep convection consumes CAPE in small-scale patches while a favorable distribution of CAPE (increasing radially outwards) is maintained on a relatively larger scale.

The void created by rising moist air is filled by the entrainment of moist air from neighboring locations and dry subsiding air in far away locations (figure \ref{fig_schema}(c)). Recently, \citet{dunkerton2009tropical} proposed the marsupial theory according to which a large pouch of moist air and large-scale circulation is established in the lower troposphere due to wave-breaking. This large-scale circulatory pattern protects the inner region of high moisture against the intrusion of dry air from regions outside this pouch. The inner region is repeatedly moistened through convection in VHTs. Further, the applicability of the marsupial theory for cyclone formation in the Bay of Bengal sea was discussed by \citet{rajasree2016appraisal}. Furthermore, a quasi-geostrophic model of moist vortex instability mechanism suggests that, a synoptic-scale vortex can form due to the coupling between the horizontal advection of the potential vorticity and the propagation of moist static energy \citep{adames2021cyclogenesis}. By virtue of moisture feedback \citep{wing2018convective}, convection fosters accumulation of moisture in neighboring regions which favors more convection, and thus contributes to intensifying the local vorticity anomalies. Furthermore, the self-aggregation of clouds through radiative-convective feedback \citep{yanai1964formation, khairoutdinov2013rotating, adames2018interactions} is known to significantly accelerate the formation of cyclones \citep{muller2018acceleration} and large-scale organization of convection \citep{holland1984dynamics, montgomery2017recent, adames2018interactions, adames2022moist, davis2015formation}. However, the mechanism of organization of localized deep convection processes is still an open question.

In light of these recent advances, understanding how local vorticity dynamics aggregate and foster organized convection at a large-scale is necessary. The various thermo-fluid processes described above are strongly related to the vorticity interactions in the flow. The large-scale background flow determines the vorticity distribution at the smaller scales. Turbulence, moisture-convection feedback and large-scale organized convection play a crucial role in determining the intensity and structure of a cyclone. 
Atmospheric turbulence modulates the intensity of cyclones, as evident in large eddy simulations of the turbulent core of cyclones \citep{oguejiofor2024role}. Further, merging of VHTs aggregates the vorticity anomalies and thus fosters a stronger and broader vortex \citep{enagonio2001tropical}. Vorticity interactions in the flow field can induce moist convection, and also be influenced by the vorticity generated through moisture feedback and deep convection. 
Therefore, we must understand the role of vorticity dynamics and the interactions at small scale, coupling of small-scale vorticity dynamics to the local convective potential and their organization at a large-scale. In this work, we study how large-scale organization \textit{emerges} from small-scale vorticity interactions. We also ask, what is the role of such large-scale organization in sustaining convection prior to the formation of a tropical depression? Can we distinguish the physical thermo-fluid processes for depressions that developing into cyclones from those that do not? To do so, we adopt the perspective of complex systems theory \citep{kauffman1995home, ladyman2013complex, ottino2003complex}. 

A complex system is one where different subsystems co-evolve through processes that are interdependent across many spatial and temporal scales \citep{ottino2003complex, siegenfeld2020introduction}. Such systems can exhibit \textit{self-organization}, i.e., spontaneous organization of interactions in the system that foster ordered dynamics and structure in the system \citep{prigogine1985self, haken2006information}. Complex systems also exhibit \textit{emergence} \citep{goldstein1999emergence, ladyman2013complex, siegenfeld2020introduction}, a phenomenon where novel coherent patterns become identifiable at the scale of the system and cannot be deduced by studying the constituents of the system independently. Emergence implies interdependence between multiple hierarchical scales of organization in a complex system \citep{goldstein1999emergence}. The Earth's climate is an archetypal complex system \citep{rind1999complexityclimate}, comprising interactions between the atmosphere, land topography and ocean, and interactions across multiple scales within the atmosphere. In a complex system, studying the emergent patterns can help infer the structure of interactions and physical processes in the system \citep{ottino2003complex, siegenfeld2020introduction, estrada2023complex}. We discuss the interdependence between the emergence of organized patterns of convection and the thermodynamic parameters of rotating moist air (diagrammatically represented in figure \ref{fig_schema}(d)).
We use the framework of complex networks to encode small-scale vorticity interactions and study the associated patterns of convection.

A network is a set of nodes (geographical locations) connected by links representing the inter-relatedness of dynamics at two locations. A recent network-based study highlighted the extent of changes in the atmospheric flow along the path of a cyclone \citep{gupta2021complex} by constructing a network of pair-wise correlations between pressure variations at different locations. Further, the various stages of cyclone mergers were classified using a novel network-based approach \citep{de2023study} where links were established based on Biot-Savart law (velocity induced between vortical perturbations between any two locations). The use of complex networks has also proven insightful for studying teleconnections in climate \citep{tsonis2006networks, tsonis2008role, donges2009complex}, and seasonal patterns of monsoon and extreme weather events \citep{wolf2020event, malik2012analysis}. Network-based approaches have offered both novel physical insights and increased predictive power by complementing conventional numerical models \citep{ludescherBoers2021network}. Here, we study vorticity interactions in the background flow during cyclogenesis using \textit{time-varying spatial proximity networks}, as described in section \ref{sec_methods}. Such networks encode time-evolving interactions between spatial neighbors alone. Due to the spatial proximity constraint, we account for sub-meso scale interactions and examine the structure of interactions on a large (synoptic) scale. We find that large-scale organized patterns emerge from small-scale vorticity interactions (section \ref{sec_result_emergence}). Examining the thermodynamic parameters of moist convection in regions of intense vorticity interactions, we find that the evolution of regions of high CAPE and high CI are strongly correlated during cyclogenesis; such correlations are absent in cases where depressions do not form cyclones (section \ref{sec_result_CAPECI}). 
Further, using modal analysis of the evolution of network connectivity (section \ref{sec_methods}), we explain the mechanism of how these large-scale patterns facilitate the propagation of local vorticity dynamics and convective processes in section \ref{sec_result_BMA}.


\section{Methods}\label{sec_methods}

\subsection{Data} 

We consider multiple depressions that formed in the north Bay of Bengal during the summer season (March- June). Some of these depressions intensify to form cyclones (developing cases, Table \ref{table_cyclonedata}), while some dissipate without reaching cyclone strength (non-developing cases, Table \ref{table_nondevdata}). The best track data is available for all the cases through the Regional Specialised Meteorological Centre (\citet{imd_besttrack}, Cyclone Warning Division, India Meteorological Department (IMD), New Delhi). The period of analysis is considered starting six days prior to the date on which a depression is identified by IMD, and ending four days after the cyclone/ non-developing depression dissipates into a weak low pressure system. The dates on which these storms form, attain peak intensity, dissipate and the period of analysis are provided in Tables \ref{table_cyclonedata} and \ref{table_nondevdata}. 
The vorticity ($\omega$) field is obtained at $850$ hPa (around $1500$ m from sea level) from ERA-5 reanalysis dataset \citet{hersbach2020era5} at multiple pressure levels, while CAPE and CI is obtained from single level ERA-5 reanalysis dataset \citet{hersbach2020era5}. We also use velocity field data for components $u,~v,~w$ at three pressure levels $825,~850, ~875$ hPa to perform vorticity budget analysis explained in section \ref{sec_meth_vortbudg}. Here, $u,~v$ are in m/s, while $w$ is provided in Pa/s as the reanalysis datasets are available in pressure coordinate system. The temporal resolution for $\omega, u,~v,~w$, CAPE and CI is hourly. Spatial resolution of $\omega$ used for network construction is $0.5^\circ \times 0.5^\circ$. 

\begin{table*}[h!]
\centering
\begin{tabular}{cccccccc}
\multirow{2}{*}{\textbf{\begin{tabular}[c]{@{}c@{}}Developing\\ case\end{tabular}}} &
  \multicolumn{4}{c}{\textbf{Dates}} &
  \multirow{2}{*}{\textbf{\begin{tabular}[c]{@{}c@{}}MSW \\ (kmph)\end{tabular}}} &
  \multirow{2}{*}{\textbf{\begin{tabular}[c]{@{}c@{}}LCP \\ (hPa)\end{tabular}}} &
  \multirow{2}{*}{\textbf{\begin{tabular}[c]{@{}c@{}}Analysis\\ period\end{tabular}}} \\
 &
  \textbf{Depression} &
  \textbf{Cyclone} &
  \textbf{PI} &
  \textbf{LPS} &
   &
   &
   \\ \hline

\begin{tabular}[c]{@{}c@{}}Bangladesh \\ (1991)\end{tabular} &
  \begin{tabular}[c]{@{}c@{}}Apr 24, \\ 12 hrs\end{tabular} &
  \begin{tabular}[c]{@{}c@{}}Apr 25, \\ 12 hrs\end{tabular} &
  \begin{tabular}[c]{@{}c@{}}Apr 29, \\ 12 hrs\end{tabular} &
  \begin{tabular}[c]{@{}c@{}}Apr 30, \\ 06 hrs\end{tabular} &
  235 &
  918 &
  \begin{tabular}[c]{@{}c@{}}Apr 18\\ to May 3\end{tabular} \\ \hline

   \begin{tabular}[c]{@{}c@{}}Nargis \\ (2008)\end{tabular} &
  \begin{tabular}[c]{@{}c@{}}Apr 27, \\ 03 hrs\end{tabular} &
  \begin{tabular}[c]{@{}c@{}}Apr 28, \\ 00 hrs\end{tabular} &
  \begin{tabular}[c]{@{}c@{}}May 2, \\ 06 hrs\end{tabular} &
  \begin{tabular}[c]{@{}c@{}}May 3, \\ 12 hrs\end{tabular} &
  167 &
  962 &
  \begin{tabular}[c]{@{}c@{}}Apr 21\\ to May 8\end{tabular} \\ \hline

     \begin{tabular}[c]{@{}c@{}}Fani \\ (2019)\end{tabular} &
  \begin{tabular}[c]{@{}c@{}}Apr 26, \\ 02 hrs\end{tabular} &
  \begin{tabular}[c]{@{}c@{}}Apr 27, \\ 06 hrs\end{tabular} &
  \begin{tabular}[c]{@{}c@{}}May 2, \\ 09 hrs\end{tabular} &
  \begin{tabular}[c]{@{}c@{}}May 4, \\ 18 hrs\end{tabular} &
  213 &
  932 &
  \begin{tabular}[c]{@{}c@{}}Apr 20\\ to May 9\end{tabular} \\ \hline
  
\begin{tabular}[c]{@{}c@{}}Amphan \\ (2020)\end{tabular} &
  \begin{tabular}[c]{@{}c@{}}May 16, \\ 00 hrs\end{tabular} &
  \begin{tabular}[c]{@{}c@{}}May 16, \\ 12 hrs\end{tabular} &
  \begin{tabular}[c]{@{}c@{}}May 18, \\ 18 hrs\end{tabular} &
  \begin{tabular}[c]{@{}c@{}}May 21, \\ 18 hrs\end{tabular} &
  240 &
  920 &
  \begin{tabular}[c]{@{}c@{}}May 10\\ to May 24\end{tabular} \\ \hline
  
\begin{tabular}[c]{@{}c@{}}Mocha \\ (2023)\end{tabular} &
  \begin{tabular}[c]{@{}c@{}}May 09, \\ 12 hrs\end{tabular} &
  \begin{tabular}[c]{@{}c@{}}May 11, \\ 00 hrs\end{tabular} &
  \begin{tabular}[c]{@{}c@{}}May 13, \\ 18 hrs\end{tabular} &
  \begin{tabular}[c]{@{}c@{}}May 15, \\ 03 hrs\end{tabular} &
  213 &
  938 &
  \begin{tabular}[c]{@{}c@{}}May 3\\ to May 19\end{tabular} \\ \hline
\end{tabular}
\caption{\label{table_cyclonedata}Information for different cyclones analyzed: Dates when a tropical depression is identified, a named cyclone is identified, the storm attains its peak intensity (PI) and when the storm dissipated to form a weak low pressure system (LPS). The 3-minute maximum sustained wind speeds (MSW) and the lowest central pressure (LCP) attained at peak intensity are listed. The time period of analysis for each of cyclones is mentioned in the last column.}
\end{table*}

\begin{table*}[h!]
\centering
\begin{tabular}{cccccccc}
\multirow{2}{*}{\textbf{\begin{tabular}[c]{@{}c@{}}Non-developing\\ case\end{tabular}}} &
  \multicolumn{2}{c}{\textbf{Dates}} &
  \multirow{2}{*}{\textbf{\begin{tabular}[c]{@{}c@{}}MSW \\(kmph)\end{tabular}}} &
  \multirow{2}{*}{\textbf{\begin{tabular}[c]{@{}c@{}}LCP \\(hPa)\end{tabular}}} &
  \multirow{2}{*}{\textbf{\begin{tabular}[c]{@{}c@{}}Analysis period\end{tabular}}} \\
 &
  \textbf{Depression} &
  \textbf{Weak LPS} &
   &
   &
   \\ \hline

\begin{tabular}[c]{@{}c@{}}1994 (D)\end{tabular} &
  \begin{tabular}[c]{@{}c@{}}Mar 21, \\ 12 hrs\end{tabular} &
  \begin{tabular}[c]{@{}c@{}}Mar 24, \\ 06 hrs\end{tabular} &
  46.3 &
  NA &
  \begin{tabular}[c]{@{}c@{}}Mar 15 to Mar 27\end{tabular} \\ \hline

   \begin{tabular}[c]{@{}c@{}}1996 (DD)\end{tabular} &
  \begin{tabular}[c]{@{}c@{}}May 7, \\ 03 hrs\end{tabular} &
  \begin{tabular}[c]{@{}c@{}}May 8, \\ 03 hrs\end{tabular} &
  55.5 &
  1000 &
  \begin{tabular}[c]{@{}c@{}}May 1 to May 11\end{tabular} \\ \hline

     \begin{tabular}[c]{@{}c@{}}2004 (DD)\end{tabular} &
  \begin{tabular}[c]{@{}c@{}}Jun 11, \\ 03 hrs\end{tabular} &
  \begin{tabular}[c]{@{}c@{}}Jun 14, \\ 09 hrs\end{tabular} &
  55.5 &
  992 &
  \begin{tabular}[c]{@{}c@{}}Jun 5 to Jun 17\end{tabular} \\ \hline
  
\begin{tabular}[c]{@{}c@{}}2022 (DD)\end{tabular} &
  \begin{tabular}[c]{@{}c@{}}Mar 3, \\ 00 hrs\end{tabular} &
  \begin{tabular}[c]{@{}c@{}}Mar 6, \\ 03 hrs\end{tabular} &
  55.5 &
  1000 &
  \begin{tabular}[c]{@{}c@{}}Feb 25 to Mar 9\end{tabular} \\ \hline

\end{tabular}
\caption{\label{table_nondevdata}Information for different non-developing cases analyzed: Dates when a tropical depression is identified, and when the storm dissipated to form a weak low pressure system (LPS). The 3-minute maximum sustained wind speeds (MSW) and the lowest central pressure (LCP) attained at peak intensity is listed. The time period of analysis for each of depression is mentioned in the last column. Here, the classification of the storm is indicated as (D) and (DD) for storms that attained maximum strength of depression and deep depression, respectively.}
\end{table*}

\begin{figure}
    \centering
    \includegraphics[width=1\linewidth]{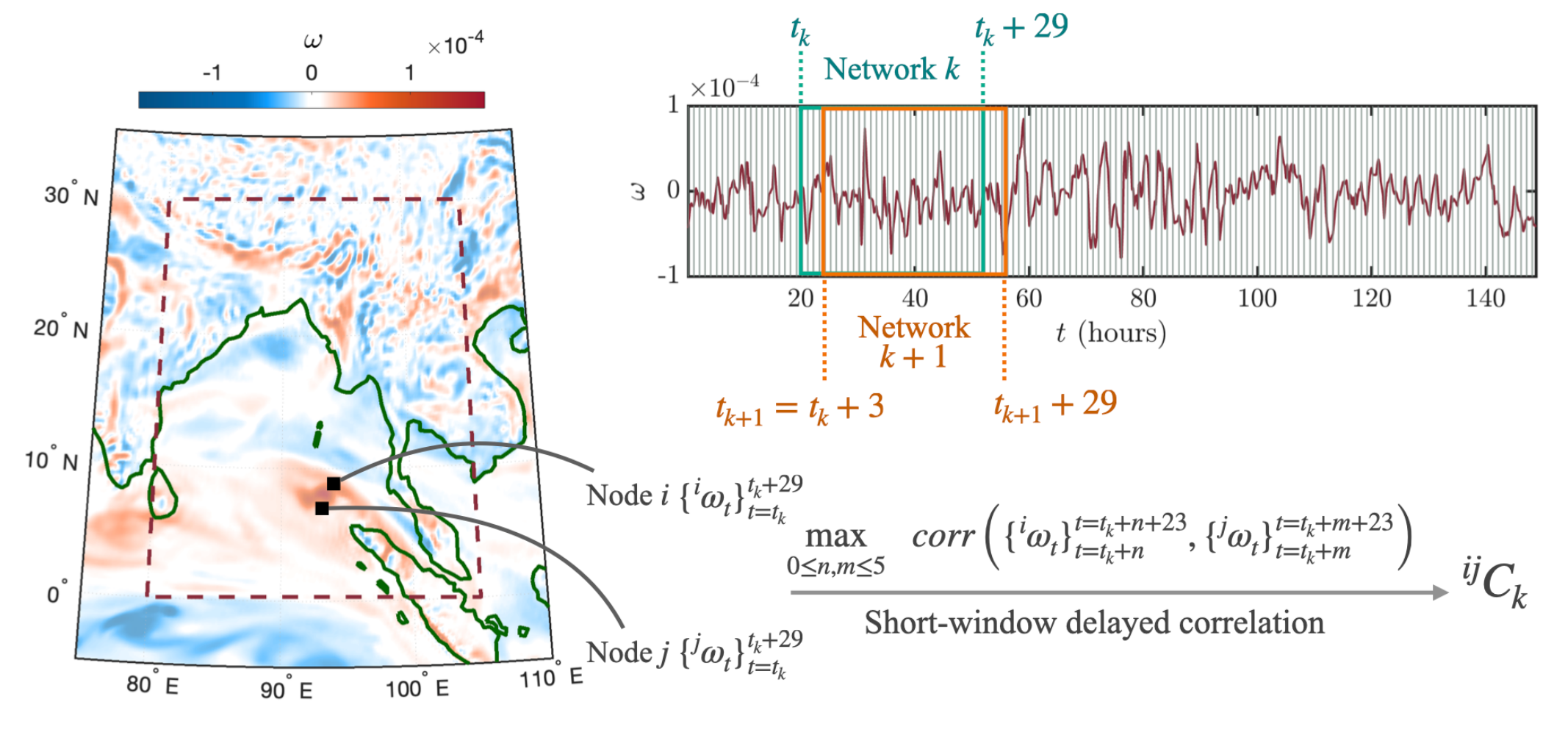}
    \caption{A schematic showing the method for the construction of time-evolving spatial-proximity networks. A network is a set of nodes connected by links; here, geographical locations are nodes and links represent statistical relation between the relative vorticity dynamics at the two nodes. Two consecutive network time frames are indicated on a representative time series plot of relative vorticity ($\omega$). Networks $k$ and $k+1$ are separated by a period of three hours and each network is constructed for a period of $29$ hours. Representative spatial variation of $\omega$ is shown in the left panel. The network is constructed in the entire spatial domain ($-5^\circ$ to $35^\circ$ N and $75^\circ$ to $110^\circ$ E) and the results from the network connectivity are analyzed in a smaller domain ($0^\circ$ to $30^\circ$ N and $80^\circ$ to $105^\circ$ E) indicated by the box drawn with red dashed line. During one network time-frame, the weight of the link between nodes $i$ and $j$ is equal to the maximum value of the short-window delayed correlation between the $29$-hour time series of relative vorticity ($\omega$) with a maximum time delay of $5$ hours. A link is established between nodes $i$ and $j$ only if the value of correlation is statistically significant ($>99$ percentile of random surrogate correlations) and both nodes are in a spatial proximity of less than $2^\circ$ in latitude and longitude. Hence, we establish links that represent local vorticity interactions between neighboring locations, and examine if patterns emerge at much larger scales.}
    \label{fig_ntwkconstruct_schema}
\end{figure}

\subsection{Time-varying spatial-proximity networks}
Geographical locations are considered as nodes in our network. The time-frame corresponding to one network is referred as a network time-frame. Each network time-frame corresponds to a 29-hour period. Two consecutive network time-frames are separated by three hours (schematic figure \ref{fig_ntwkconstruct_schema}). Networks evolve in steps of three hours to help capture the dynamics during formation and intensification of cyclones that occurs generally over a period of six to twelve hours \citep{gray1998formationrev}.

A connection is established between any two nodes during a network time-frame depending on two conditions: \textbf{(I)} Two nodes $i$ and $j$ are connected if these nodes are in spatial proximity of each other within $\leq 2^\circ$ in latitude or longitude. The $2^\circ$ spatial constraint ($\approx 220$ km) implies that the network encodes local meso and sub-meso scale interactions. \textbf{(II)} A link is established between the two nodes if the relative vorticity dynamics at the two locations evolve similarly over a short time period, as determined using statistically significant short-window delayed correlations.

The time duration corresponding to the $k^{\text{th}}$ network time-frame is $t\in[t_0+3(k-1)+1,t_0+3(k-1)+29]$, where $t_0+1$ denotes the starting time instant of the time period of analysis (refer Tables \ref{table_cyclonedata} and \ref{table_nondevdata}). Then, for a node $i$, the time series of relative vorticity during a network time-frame $k$ is denoted as $\{^i\omega_t\}_{t=t_k}^{t_k+28}$, where $t_k=t_0+3(k-1)+1$. The value of short-window delayed correlations for the time series $\{^i\omega_t\}_{t=t_k}^{t_k+28}$ and $\{^j\omega_t\}_{t=t_k}^{t_k+28}$ at nodes $i$ and $j$ is given by Eq. \ref{eqn_maxcorr}. Correlations are calculated between the 24-hour time series of $^i\omega$ and $^j\omega$ allowing a maximum time-delay of five hours within the network time-frame $k$. 
\begin{equation}\label{eqn_maxcorr}
   ^{ij} C_k
    =\underset{0 \leq n,m \leq 5}{\max}\;\;corr\left(\{^i\omega_t\}_{t=t_k+n}^{t=t_k+n+23},\{^j\omega_t\}_{t=t_k+m}^{t=t_k+m+23}\right)
\end{equation}
Adjacency matrix $A_k$ encodes the connections during the network time-frame $k$. The weight of the connection between nodes $i$ and $j$ is equal to the maximum short-window delayed correlation; hence, $A_k[i,j]=A_k[j,i]=~^{ij}C_k$. Since correlation is a symmetric metric, the adjacency matrix is symmetric and connections between the nodes are undirected. 
To enable network construction on GPU and accelerate computation, we compute correlations in the Fourier domain; time-delayed correlation is essentially a convolution in the time domain and a product of the Fourier transforms in the frequency domain. Further, we calculate the node strength of node $i$ as $NS_i=\Sigma_{j}A_k[i,j]$, where $j$ is a spatial neighbor of $i$. Thus the node strength represents the network connectivity and the overall strength of coherence in the vorticity dynamics between node $i$ and its neighbors. 

The spatial extent for network analysis of each cyclone is chosen over the BoB such that it spans the entire path of the cyclone and covers the region of large-scale circulations associated with it. Spatial domain of network construction is between $-5^\circ$ to $35^\circ$ N and $75^\circ$ to $110^\circ$ E. Since we construct a spatial proximity networks, only those nodes that are close to the boundaries can possibly be influenced by locations outside the domain of analysis. Hence, after constructing the network in a larger domain, we consider the results in a restricted domain in $0^\circ$ to $30^\circ$ N and $80^\circ$ to $105^\circ$ E (dotted box in figure \ref{fig_ntwkconstruct_schema}, to avoid the boundary effects. Thus, the spatial extent is $30^\circ \times 25^\circ$ in latitude and longitude corresponding to a length-scale of $\approx 3100$ km and $\approx2778$ km, respectively (km/per degree longitude varies with latitude, from 96 km to 111 km between $0-30^\circ$ latitude. The length-scale is estimated by taking an average of this range and multiplying with $30^\circ$). In comparison, the average gale force wind radius (i.e. radius of MSW of 17 m/s = 61.2 kmph) of a super cyclonic storm in BoB is $\approx240$ km \citep{imd_sop}. Thus, the area of analysis is approximately $40$ times the area covered by the vortex of a super cyclonic storm. 
In \textcolor{black}{Supplementary S1}, we show that choosing a smaller spatial proximity constraint, such as  $1.5^\circ$ or $1.75^\circ$, does not change the emergent patterns and our inferences remain similar. Intricacies of the emergent pattern are enhanced on using higher resolution ($0.25^\circ \times 0.25^\circ$) vorticity data for network construction (\textcolor{black}{Supplementary S2}).

\subsection{Broadcast mode analysis}\label{sec_meth_BMA}

In order to understand how large-scale patterns of organized convection emerge, we study the propagation of local connectivity in flow perturbations. We use broadcast mode analysis \citep{katz1953new, grindrod2014dynamical, yeh2021network} that measures the importance of nodes in propagating information. Using this method, we track how a perturbation at one node is communicated across different nodes (spatially) and over time. 

A perturbation $f$ originating at node $i$ in a network during a fixed time-frame $k$ can propagate to neighbors along paths of varying link lengths. The strength of communication from node $i$ to its immediate neighbor is proportional to the weight of the connection between the two nodes. Thus, such a propagation can be denoted as $\alpha A_k f$, where $\alpha$ denotes a scaling (or attenuation) factor, called the walk-downweighting parameter, along the path between two nodes. Similarly, the perturbation can propagate further to second nearest neighbors with strength $(\alpha A_k)^2 f$. Then, the propagation of $f$ along paths of multiple lengths is $(I+\alpha A+\alpha^2 A^2+\alpha^3 A^3+....)f=(I-\alpha A)^{-1}f$; here, $(I-\alpha A)^{-1}$ is the Katz function for one network time-frame, \citep{katz1953new}.
We can consider a composite of Katz functions for several network time-frames representing the propagation of perturbations due to the time-evolution of network connectivity. This composite Katz function is called the communicability matrix ($S_m$), and is evaluated for network time-frames $k_1$ to $m$ as given in Eq. \ref{eq_commmatrix}.
\begin{equation}\label{eq_commmatrix}
    S_m=(I-\alpha A_m)^{-1}.... (I-\alpha A_{k_1+1})^{-1}(I-\alpha A_{k_1})^{-1}
\end{equation}

The ability of any node $i$ to communicate to $j$ depends on $\textbf{U}(t_m)=S_m+I$. Thus, $\textbf{U}(t_m)$ denotes a dynamical system in which the perturbations propagate across nodes over time. We can now extract two modes from the matrix $\textbf{U}(t_m)$, namely, (i) the dynamic broadcast mode $\textbf{b}=[b]=\textbf{U}\textbf{1}$ (sum of rows of $\textbf{U}$) and (ii) the dynamic receiving mode $\textbf{r}=[r]=\textbf{U}^T\textbf{1}$ (sum of columns of $\textbf{U}$) \citep{grindrod2014dynamical}. Nodes with higher values of $b$ (or $r$) can broadcast (or receive) information of perturbations with higher propensity. Thus, the broadcast and receiving modes highlight nodes that are most influential and most influenced over several network time-frames, respectively. In our analysis, we consider ten network time-frames together, which correspond to $56$-hours of data of the relative vorticity field. Also, $\alpha=0.03$; this value is chosen as less than the spectral radius of all the adjacency matrices of the time-varying complex networks, following the criterion given in \citep{grindrod2014dynamical}. Small variations in the value of $\alpha$ below the threshold determined using this criterion does not alter the results.

\subsection{Vorticity budget analysis}\label{sec_meth_vortbudg}

Equation \ref{eq_vortbudgetfull} describes the evolution of absolute vorticity \citet{pedlosky1987geophysical, holton2012introduction}. The first term on the right hand side is the tilting term, the second term is the stretching (or divergence) term, the third term represent baroclinic vorticity generation, and the fourth term represents dissipation of vorticity due to friction. Vertical vorticity is expressed as sum of relative vorticity and planetary vorticity (Eq. \ref{eq_vortbudg_EXTRA}). 

\begin{equation}\label{eq_vortbudgetfull}
\frac{D\vec{\omega}_a}{Dt} = (\vec{\omega}_a \cdot \nabla)\vec{V} - \vec{\omega}_a(\nabla \cdot \vec{V}) + \frac{1}{\rho^2}\nabla\rho \times \nabla p + \nabla \times \vec{F}
\end{equation}

\begin{equation}\label{eq_vortbudg_EXTRA}
\vec{\omega}_a = \nabla \times \vec{V} + 2\vec{\Omega}
\end{equation}
Let us consider only the vertical component of the equation, $\eta = \vec{\omega}_a \cdot \hat{k} = \zeta + f$ where, $\zeta = \frac{\partial v}{\partial x} - \frac{\partial u}{\partial y}$ and $f = 2\Omega \sin \phi$. We consider isobaric pressure coordinates; vertical wind is expressed in $\text{Pa}/\text{s}$ and $\frac{\partial }{\partial z}$ is converted to $\frac{\partial }{\partial p}$ in the pressure coordinate. The solenoidal term is neglected on each constant pressure surface \citep{kumar2025unravelling}. As a result, we obtain Eq. \ref{eq_vortbudget_FINAL}. 

\begin{equation}\label{eq_vortbudget_FINAL}
\frac{\partial \zeta}{\partial t} = \underbrace{-(\zeta + f)\left(\frac{\partial u}{\partial x} + \frac{\partial v}{\partial y}\right)}_{{S_\zeta}} 
\underbrace{- w\frac{\partial \zeta}{\partial p}}_{{VA}} 
~~~\underbrace{- \left(u\frac{\partial \eta}{\partial x} + v\frac{\partial \eta}{\partial y}\right)}_{HA} 
\underbrace{- \left(\frac{\partial w}{\partial x}\frac{\partial v}{\partial p} - \frac{\partial w}{\partial y}\frac{\partial u}{\partial p}\right)}_{T_\zeta} 
+ R
\end{equation}

Here, $w$ is in Pa/s, $u,~v$ are in m/s, and $R$ is the residual term containing primarily of frictional effects. To aid our understanding from the modal analysis of network connectivity, we perform vorticity budget analysis at $850$ hPa using velocity data at $825,~850,~875$ hPa. The \textit{tilting} term, $T_\zeta$, represents redistribution of existing vorticity from one spatial plane into another. Significant tilting in spatial pockets implies conversion of horizontal vorticity to vertical vorticity; positive values imply that the vertical vorticity is cyclonic (anticlockwise), while negative values imply that the vertical vorticity is anticyclonic (clockwise). Further, the stretching term $S_\zeta$ represents amplification or reduction in vertical vorticity due to convergence or divergence, respectively. Positive stretching implies amplification of local vertical vorticity through low-level convergence contributing to cyclonic spin-up. Similarly, positive vertical advection ($VA$) implies strong updrafts transporting vertical vorticity to higher altitudes. Positive horizontal advection ($HA$) represents regions where vertical vorticity is transported horizontally at the same pressure level. Such positive horizontal advection is expected to be particularly dominant along bands of entrainment close to the cyclone core where moist wind is entrained inwards by the spinning cyclonic system. 


\section{Results}
\subsection{Emergence of large-scale patterns from small-scale vorticity interactions}\label{sec_result_emergence}

Let us consider the specific example of cyclone Amphan. On May 13 (2020), an identifiable low pressure area formed as a result of easterly wave perturbations in the southeast Bay of Bengal (BoB) region. This low pressure perturbation formed a depression on the midnight of May 16 and rapidly intensified into a deep depression, and subsequently into cyclone Amphan on the afternoon of May 16 \citep{imd2020amphanpr, bhowmick2020cycloneamphan, singh2024intensificationAmphan}. We construct time-varying spatial proximity networks from the relative vorticity correlations at $850$ hPa starting from May 10 (2020), as discussed in section \ref{sec_methods}. 
A weighted link is established between two nodes (locations), with the weight equal to the value of the maximum delayed correlation between the vorticity dynamics at the two locations (Eq. \ref{eqn_maxcorr}) for a 29-hour period allowing a maximum delay of five hours. Also, nodes are connected only if they are located spatially close to each other within sub-meso length scales. Consecutive networks are separated by a period of three hours.
We visualize the strength of network connectivity at each location, called the node strength ($NS$), by summing up the weights of links (correlations) between a node and its geographical neighbors. A patch of high node strength highlights regions where the relative vorticity dynamics is significantly correlated among neighboring locations. Thus, a patch of high node strength highlights pockets of `spatially coherent vorticity dynamics'.

\begin{figure}
\centering
\includegraphics[width=1\linewidth]{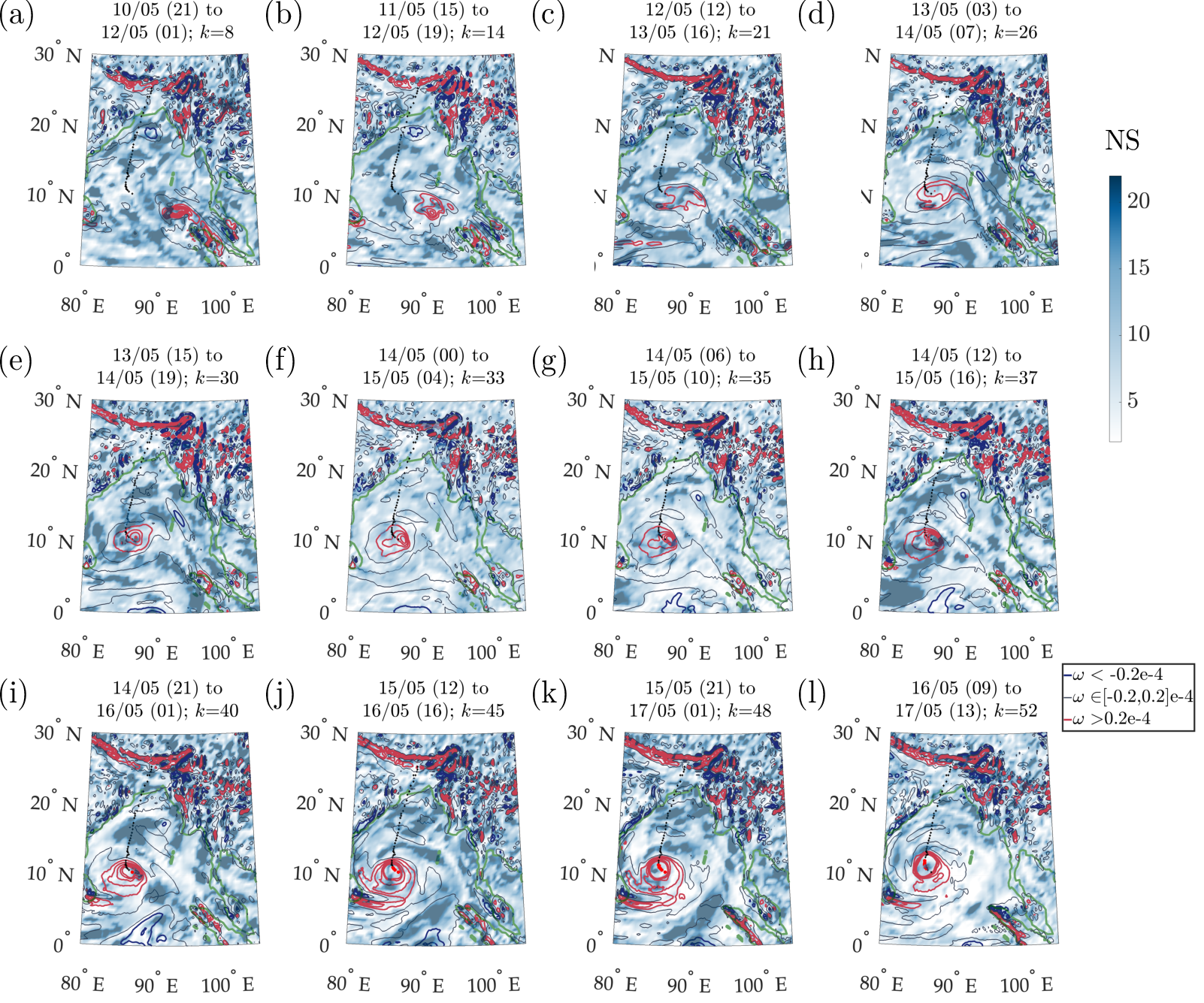}
    \caption{(a-l) Spatial distribution of node strengths $NS$ obtained from the network analysis \textbf{prior to the formation} of tropical cyclone Amphan (2020) during time-frame $k$, overlapped with contour lines representing isolines of relative vorticity ($\omega$) averaged over 29-hour period during the time-frame $k$. The time periods and network frame number ($k$) are mentioned in the title of each subplot (date/month (hrs) format). Strong positive and negative $\omega$ are represented by red and blue lines respectively, while black contours represent regions of weak vorticity. Patches of high node strengths appear to be arranged in a large-scale pattern that evolves with $k$. The trajectory of cyclone Amphan (2020) is represented by black dots beginning from the tropical depression state identified by IMD on 16/05 (00 hrs). Larger red dots along the trajectory represent the location of cyclone Amphan during that time-frame. For visualizing the evolution of the large-scale patterns over consecutive network time-frames, refer to the Movies 1 and 2, refer Supplementary S4.
  }
    \label{fig_NSvort_formn}
\end{figure}

On visualizing the spatial distribution of $NS$ (figure \ref{fig_NSvort_formn}), we discover that large-scale organized patterns emerge from small-scale vorticity interactions. Much prior to the formation of a depression, several patches of coherent vorticity dynamics appear independently and are spread across in the Bay of Bengal (BoB) region (figure \ref{fig_NSvort_formn}(a)). Several such patches emerge and dissipate over the period of next two days in the central and northern BoB region close to the coast (figure \ref{fig_NSvort_formn}(b-d)). Some of these patches emerge in regions far from the location of the low pressure system, and in regions of relatively weak vorticity perturbations associated with the background flow. The term `weak vorticity' should not be confused with negligible vorticity fluctuations; instead the term here refers to the vorticity perturbations that are weak relative to the strength of the much stronger vorticity around the low pressure anomaly of the incipient vortex. 

As the low pressure anomaly develops into an identifiable low pressure system, some of the small patches of high node strengths emerge adjacent to each other forming larger extended patches that appear along circular arcs, for example in figure \ref{fig_NSvort_formn}(c,e,h,j). Such extended structures emerge spontaneously prior to the formation of an identifiable depression. As the low pressure system further intensifies to form a tropical depression and subsequently a cyclone, we observe some of the patches of high node strengths occur in regions of high positive vorticity, whereas some patches occur in north BoB region in areas of relatively weak vorticity dynamics (figure \ref{fig_NSvort_formn}(j-l)). Further, as the cyclone intensifies, patches of coherent vorticity dynamics occur predominantly in regions of high positive vorticity associated with the cyclone and also in the wake of the cyclone (Supplementary S4). 

Interestingly, even though by construction, the network encodes local correlations in vorticity dynamics, the patches of coherent vorticity dynamics appear to be spontaneously organized in a synoptic-scale pattern. 
The spatially distributed strength of network connectivity appears to increase and decrease repeatedly with time; for example, compare figure \ref{fig_NSvort_formn}(e,f), or figure \ref{fig_NSvort_formn}(j,l). The spatially averaged node strength ($\langle NS \rangle$) exhibits oscillations of diurnal timescale. Such oscillations occur much prior to the formation of a depression, while the spatially averaged vorticity ($\langle \omega\rangle$) begins increasing only during the intensification of depression into a cyclone (figure \ref{fig_TS_ntwk}(a)). Similar emergence of large-scale patterns and diurnal oscillations in network connectivity are obtained for other examples of cyclone-forming depressions (\textcolor{black}{refer Supplementary Movies, and S4}).

On observing the node strength distribution of each network time-frame consecutively (Supplementary S4), we find that some patches remain consistent across several consecutive networks and appear to shift in location. Even though some of the initially generated patches dissipate, we find that new patches appear in adjacent locations over subsequent network time-frames, and the cyclonic motion of the large-scale pattern becomes evident (refer Supplementary Movies for such visualization). Such patches align into structures that appear to co-rotate in anti-clockwise direction (cyclonic sense of rotation), even prior to the formation of the tropical depression. Note that, even though some of the patches of high node strengths appear to `move', these results do not imply material transport of fluid between consecutive time-frames. Instead, we interpret the motion of such patches as the percolation of spatial correlations of vorticity dynamics from the region of high correlations to its neighboring regions (plausible mechanism of such percolation of correlations is discussed in subsequent sections). 

Further, we note that the dynamics of the patches of spatially coherent vorticity correlations resembles the evolution of vortical hot towers (VHTs) as discussed by \citet{montgomery2006vortical, montgomery2017recent}. VHTs are associated with high vorticity anomalies, and they intensify the local vorticity fluctuations. Similar to the distribution of VHTs \citep{montgomery2006vortical}, we find that an initially disorganized distribution of coherent vorticity patches becomes organized close to cyclogenesis. Several patches appear to merge and organize into extended structures similar to the aggregation of VHTs prior to the formation of a large vortex. However note that, the patches cannot represent the precise location of VHTs but instead encode the local interactions due to the vorticity fluctuations in the flow. In section \ref{sec_result_BMA}, we exploit the evolution of the network connectivity to unravel how patches of locally coherent vorticity dynamics become organized at a large-scale. In the next section, we examine how convective activity associated with these patches evolves during cyclogenesis.


\subsection{Interdependence of convection potential and inhibition on organization of coherent vorticity dynamics}\label{sec_result_CAPECI}

Given the organization of patches of coherent vorticity dynamics into large-scale patterns, we ask if moist convection aids and is influenced by such self-organization. We examine convective available potential energy (CAPE) and convective inhibition (CI) associated with the patches of high $NS$, i.e., patches of coherent vorticity dynamics. 
Let us define $t_{cr}$ as the time at which a depression was identified. We identify regions of intense vorticity interactions by setting a threshold on $NS$. Further, let $\langle CAPE \rangle _{HNS}$ denote the average of CAPE in locations that exhibit high node strengths during a fixed time-frame of the network analysis. We also define the \textit{fractional intensity} of CAPE as the ratio of average CAPE in regions of high node strengths with respect to the average CAPE over the entire region of analysis ($\langle CAPE \rangle _{HNS}/\langle CAPE\rangle$). Such fractional intensity of CAPE (or CI) helps quantify the convection potential (or inhibition) associated with spatially coherent vorticity dynamics.

\begin{figure}
    \centering
    \includegraphics[width=0.8\linewidth]{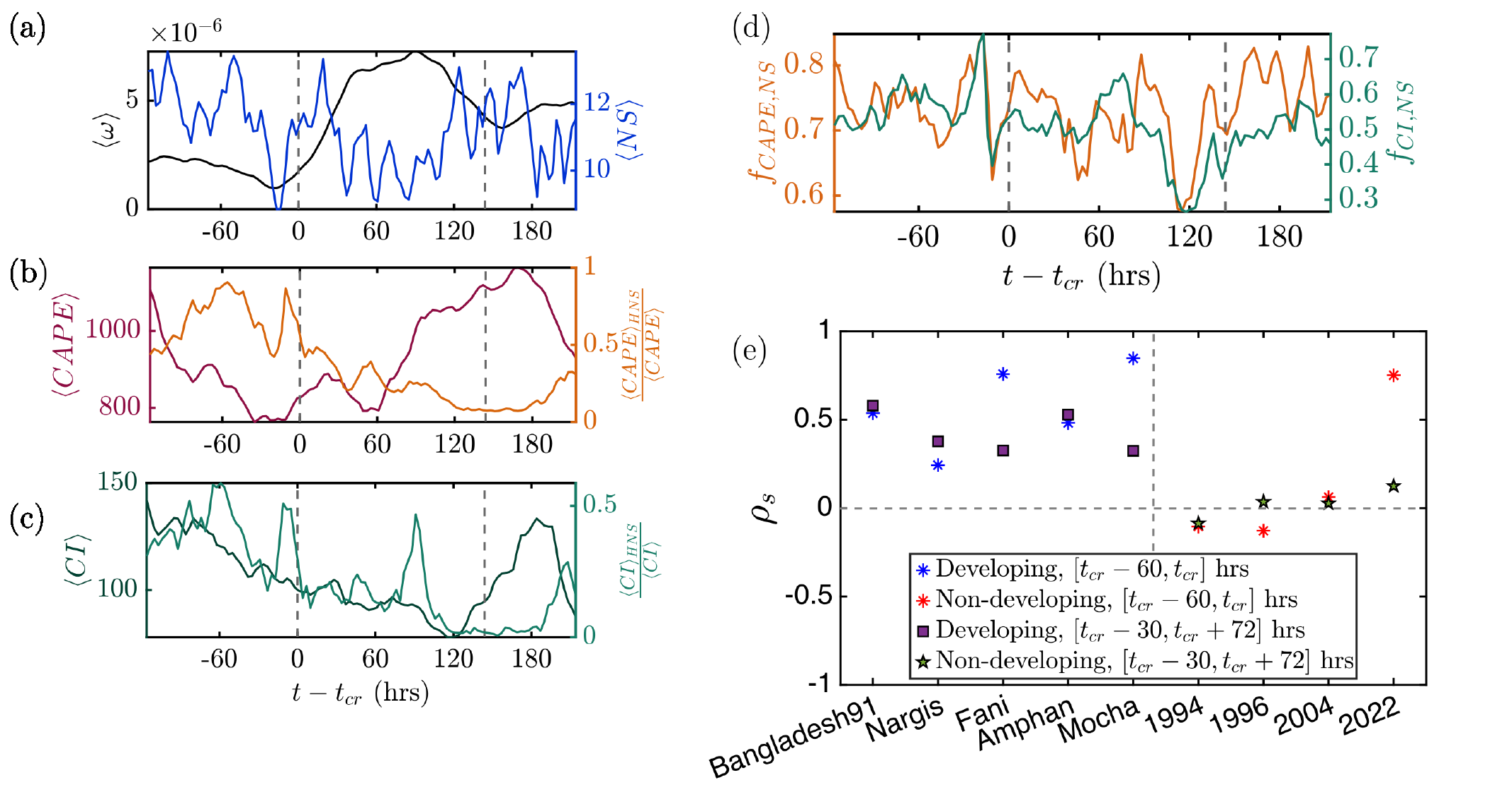}
    \caption{(a) Temporal variation of the spatial average of vorticity strength ($\langle\omega\rangle$) and node strength ($\langle NS \rangle$) during each network time-frame. (b) Overall spatially averaged intensity ($\langle CAPE\rangle$), and fractional intensity ( $\langle CAPE \rangle _{HNS}/\langle CAPE\rangle$) of convection potential. (c) Overall spatially averaged intensity ($\langle CI\rangle$), and fractional intensity of convection inhibition ( $\langle CI \rangle _{HNS}/\langle CI\rangle$). Subscript $HNS$ represents spatial average over regions of high node strengths where $NS>15$. (d) Variations of the number density of locations, $f_{CAPE,NS}$ where $NS>15$ and $CAPE>900$ J/kg, and $f_{CI,NS}$ where $NS>15$ and $CI>200$ J/kg. Subplots (a-d) are for cyclone Amphan; abscissa demarcates time relative to the critical time $t_{cr}$ at which a depression is identified (indicated by a dashed line at zero). The second dashed line indicates the time when the cyclone intensity began weakening. (e) Spearman rank correlation ($\rho_s$) between $f_{CAPE,NS}$ and $f_{CI,NS}$ for various developing and non-developing cases during time periods, (i) $P_1=[t_{cr}-60, t_{cr}]$ hrs, and (ii) $P_2=[t_{cr}-30, t_{cr}+72]$ hrs.}
    \label{fig_TS_ntwk}
\end{figure}

\begin{figure}
    \centering
    \includegraphics[width=0.9\linewidth]{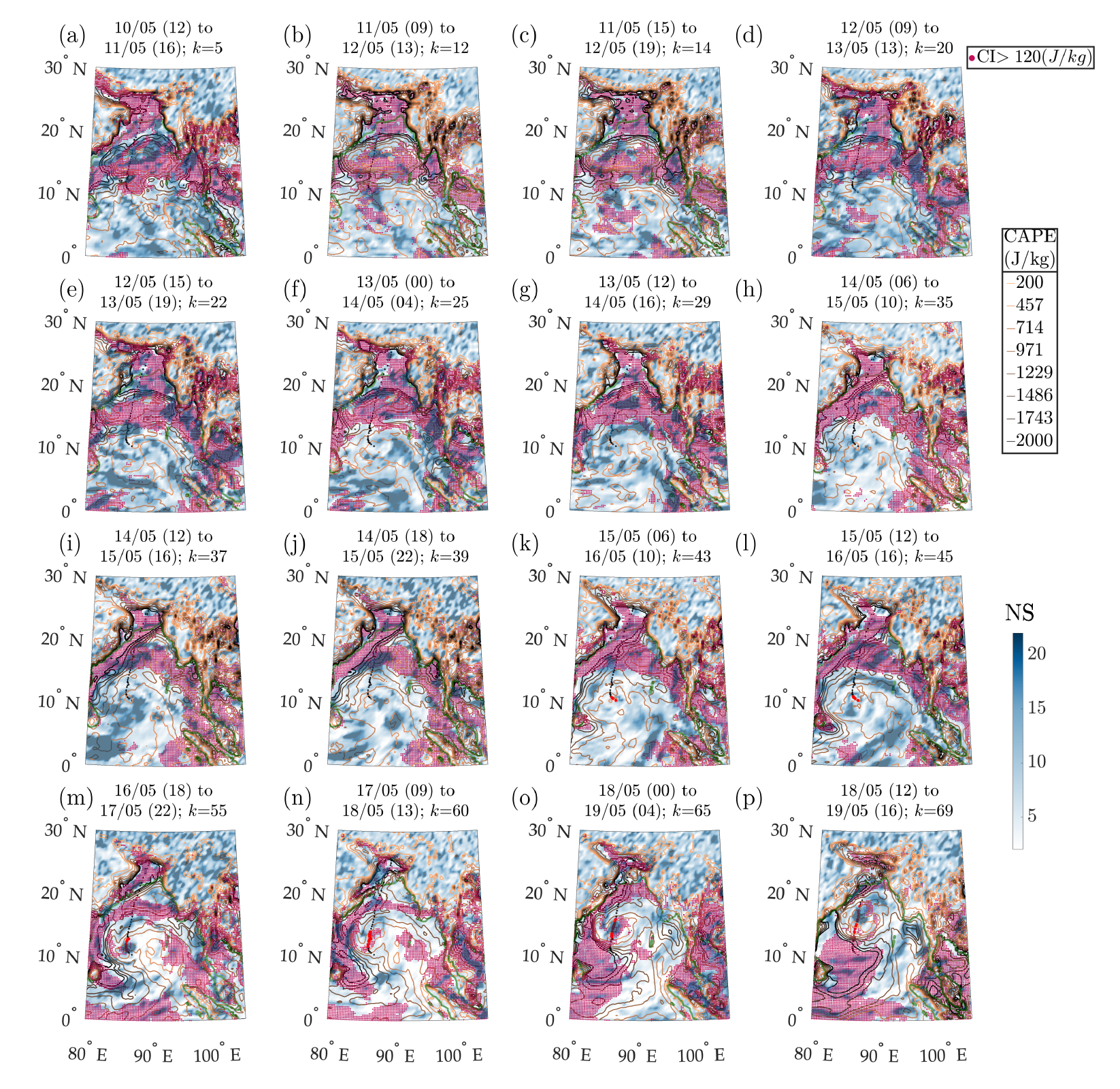}
    \caption{Simultaneous evolution of node strengths ($NS$) obtained from vorticity correlation networks, and 29-hour average of convective available potential energy (CAPE, contours) and regions of strong convective inhibition (CI, marked by magenta dots) during the (a-j) formation of an identifiable tropical depression, and (k-p) its intensification to form cyclone Amphan and subsequent dissipation. The 29-hour average of CAPE and CI are taken corresponding to the network time-frame indicated in each subplot (date/month (hrs) format). Also, the spatial contours of CAPE do not vary much during cyclogenesis, evident in subplots (a-l). Initially, regions of strong CI are spread in northern and central BoB (see patches of red dots in (a-d)). Also, CI reduces in central BoB during cyclogenesis (compare (a-d) to (e-l)), and increases in the wake of the cyclone during its intensification (m-p). Some patches of high node strength are associated with high CAPE and some with high CI during each network time-frame. Some patches of coherent vorticity dynamics initially appear near the boundaries between strong and weak CI; for example, see patches in northern and central BoB region in (a-c). Subsequently, we observe that CI weakens in these regions where CI and $NS$ were simultaneously high (compare CI along patches around $10^\circ$N and $85^\circ$E in central BoB region in (a-e)).}
    \label{fig_capeci_all}
\end{figure}

For cyclone Amphan, $\langle CAPE\rangle$ shown in figure \ref{fig_TS_ntwk}(b) initially decreases and remains constant prior to the formation of a depression; $\langle CAPE\rangle$ starts increasing prior to but very close to $t_{cr}$ and the increase becomes more significant after cyclone Amphan intensified (around $t_{cr}+60$ hrs). Spatial distribution of CAPE shown in figure \ref{fig_capeci_all}(a-l) varies minimally during cyclogenesis. On the other hand, the fractional intensity of CAPE shown in figure \ref{fig_TS_ntwk}(b) increases noticeably more than $60$ hrs prior to the formation of a depression (notice the peak at $-60$ hrs in $\langle CAPE \rangle _{HNS}/\langle CAPE\rangle$ in figure \ref{fig_TS_ntwk}(b)). Evidently, even though the spatial average of CAPE does not vary significantly, convective potential increases in the patches of coherent vorticity dynamics. For other examples shown in Supplementary S3, we find that total CAPE in the environment may decrease or increase close to the formation of a depression. However, the fractional intensity of CAPE in regions of locally coherent vorticity dynamics increases observably for the various developing cases, but not for the non-developing cases. 
This result depicts that high CAPE is associated with regions of locally coherent vorticity dynamics. As stated by \citet{dunkerton2009tropical}, repeated moistening occurs locally and fosters deep convection across several locations within a large-scale circulation prior to cyclogenesis. However, how such repeated moistening occurs and becomes organized at large-scales is yet to be understood. Here, we discover that a large fraction of the total CAPE in the synoptic region is associated with patches of locally coherent vorticity dynamics prior to the formation of a depression, and can foster deep convection subsequently. These patches with high CAPE can become organized into large-scale patterns eventually, as discussed earlier (figure \ref{fig_NSvort_formn}). 

On the other hand, convective inhibition (CI) weakens in the central BoB during the genesis of cyclone Amphan, and the spatial spread of CI decreases over time (compare figure \ref{fig_capeci_all}(a-d) to (e-l)). As the cyclone intensifies, regions of significant CI develop in the wake owing to subsidence of dry air in figure \ref{fig_capeci_all}(m-p). The spatially averaged intensity of convection inhibition ($\langle CI\rangle$ in figure \ref{fig_TS_ntwk}(c)) starts decreasing much prior to the formation of the depression, and decreases further as the depression forms a cyclone and as the cyclone intensifies. In spatial pockets where node strengths and CI are high simultaneously during initial time-frames (refer figure \ref{fig_capeci_all}(a-c)), we find that CI decreases in these regions during subsequent time-frames (compare figure \ref{fig_capeci_all}(a-c) with (d,e)).
Further, more patches of high node strengths develop in the central BoB region after the weakening of CI (e.g., figure \ref{fig_capeci_all}(e-g)). 
$\langle CI\rangle$ starts increasing almost $30$ hrs prior to when the cyclone starts dissipating (figure \ref{fig_TS_ntwk}(c)). Further, the fractional intensity of CI in regions of high node strengths ($\langle CI \rangle _{HNS}/\langle CI\rangle$, figure \ref{fig_TS_ntwk}(c)) exhibits a generally decreasing trend along with significant deviations prior to the formation of a depression. For instance, notice the increase in $\langle CI \rangle _{HNS}/\langle CI\rangle$ at $-10$ hrs and $-60$ hrs. Further $\langle CI \rangle _{HNS}/\langle CI\rangle$ spikes at $+90$ hrs in figure \ref{fig_TS_ntwk}(c), i.e., thirty hours before $\langle CI\rangle$ starts increasing. Clearly, some of the regions of coherent vorticity dynamics are associated with strong convection inhibition prior to $t_{cr}$, and also prior to the dissipation of the cyclone. Such peaks in the fractional intensity of CI can be associated with increasing downdrafts in far away regions, or with patches where CAPE has been consumed locally due to deep convection, such as via VHTs.


To further evaluate the spatial spread of regions of coherent vorticity dynamics associated with high CAPE and CI, we find the number density of locations, $f_{CAPE,NS}$ and $f_{CI,NS}$ (figure \ref{fig_TS_ntwk}(d)). The number density $f_{CAPE,NS}$ is calculated as the \textit{number} of nodes (locations) where both CAPE and node strength ($NS$) are simultaneously high during a network time-frame, normalized by the total number of nodes with high node strengths. Notice, $f_{CAPE,NS}$ and $f_{CI,NS}$ oscillate almost synchronously just before and after the formation of the depression for cyclone Amphan (other developing and non-developing cases are shown in Supplementary S3). Interestingly, the spatial spread (number density) of convective activity appears to dominate the role of convection intensity during cyclogenesis, as discussed in \citet{davis2015formation}. We find the Spearman rank correlation between $f_{CAPE,NS}$ and $f_{CI,NS}$ during two periods: (i) prior to the formation of a depression ($P_1=[t_{cr}-60, t_{cr}]$, corresponding to two and a half days before a depression was identified) and (ii) during the formation and intensification of the depression ($P_2=[t_{cr}-30, t_{cr}+72]$, corresponding to a period of one day before and 3 days after the formation of a depression), for various developing and non-developing cases.

We discover that the correlation of $f_{CAPE,NS}$ and $f_{CI,NS}$ is positive and significant for several cases where a depression intensifies to form a cyclone during both periods $P_1$ and $P_2$, but negligible for cases when a depression dissipates without forming a cyclone (figure \ref{fig_TS_ntwk}(e)). For a non-developing case of depression formed in 2022, we see that such correlation is positive during the time period $P_1$ but decreases drastically and becomes negligible during period $P_2$; possibly indicating that conditions for cyclogenesis were not suitably sustained. Thus, the correlation between convective processes occurring at various locations is significant and sustained throughout the duration of intensification of a tropical perturbation into a depression and further into a cyclone. For non-developing cases, such correlation is not sustained during either the formation or the intensification of a depression. These results imply that the large-scale pattern of local vorticity interactions aid in sustaining organized moist convection for cyclogenesis. Further, the organized patterns of locally coherent vorticity dynamics induce a coupling between CAPE and CI at different locations in the domain. Hence, the organization of coherent vorticity dynamics plays a significant role in organizing the small-scale moist convection processes during cyclogenesis; this role is investigated further in the next section.


\subsection{Large-scale modes of small-scale vorticity interactions}\label{sec_result_BMA}
\begin{figure}
    \centering
    \includegraphics[width=1\linewidth]{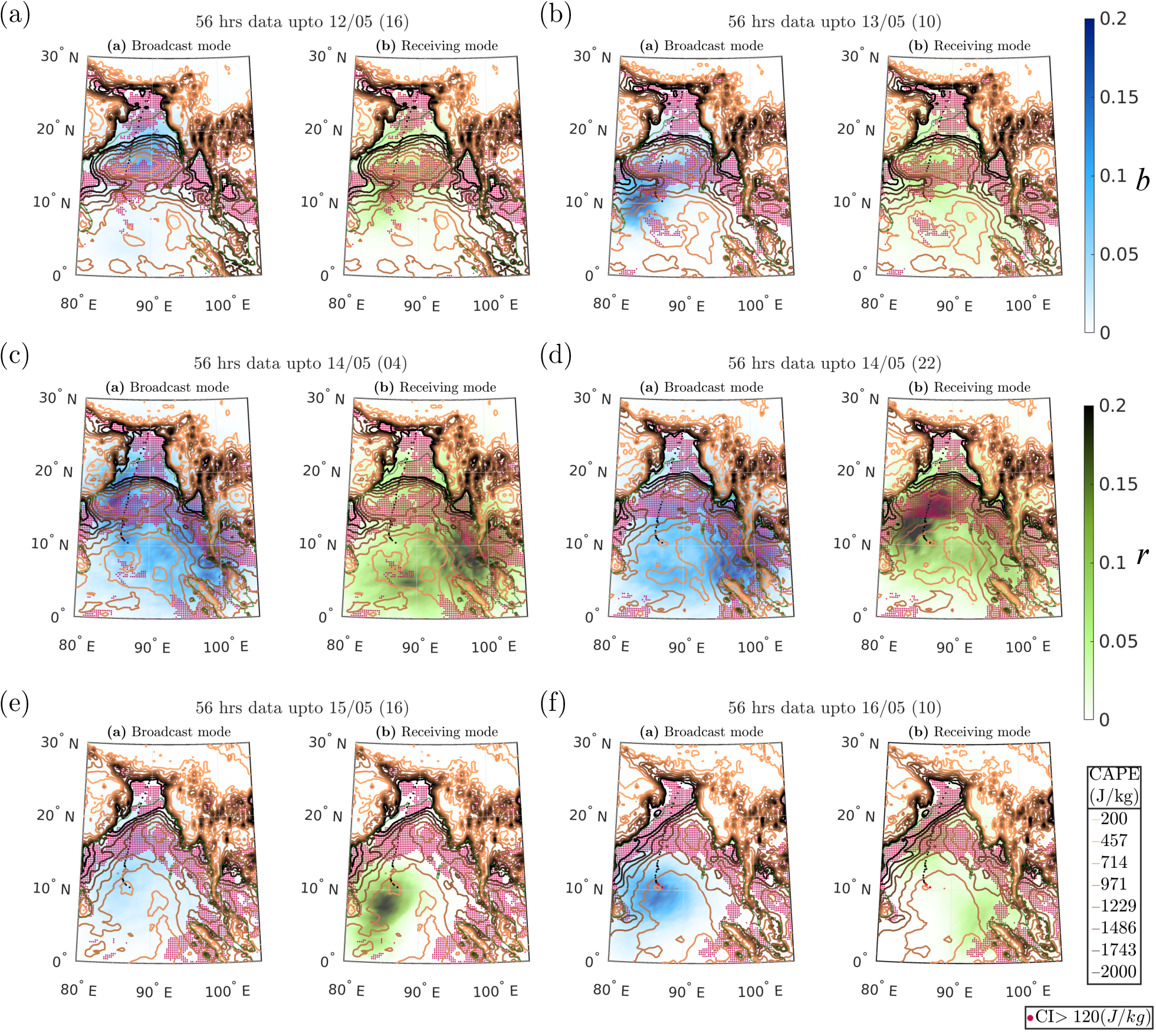}
    \caption{Spatial distribution of broadcast ($b$) and receiving ($r$) modes during different periods prior to the formation of tropical cyclone Amphan (a-f). These modes are obtained from modal analysis of $10$ networks taken together corresponding to $56$-hour period. The times are mentioned in date/month (hrs) format. The corresponding time period is mentioned above each subplot. Overlapping are the 56-hour average of convective available potential energy (CAPE, contours) and regions of strong convective inhibition (CI, marked by red dots). The broadcast mode highlights regions from where propagation of flow perturbations is most influential, and the receiving mode highlights regions where reception of the influence of flow perturbations is highest. The broadcast and receiving modes reveal large-scale communication of local coherence in vorticity dynamics during cyclogenesis.}
    \label{fig_BMA_Amphan_genesis}
\end{figure}

\begin{figure}
    \centering
    \includegraphics[width=1\linewidth]{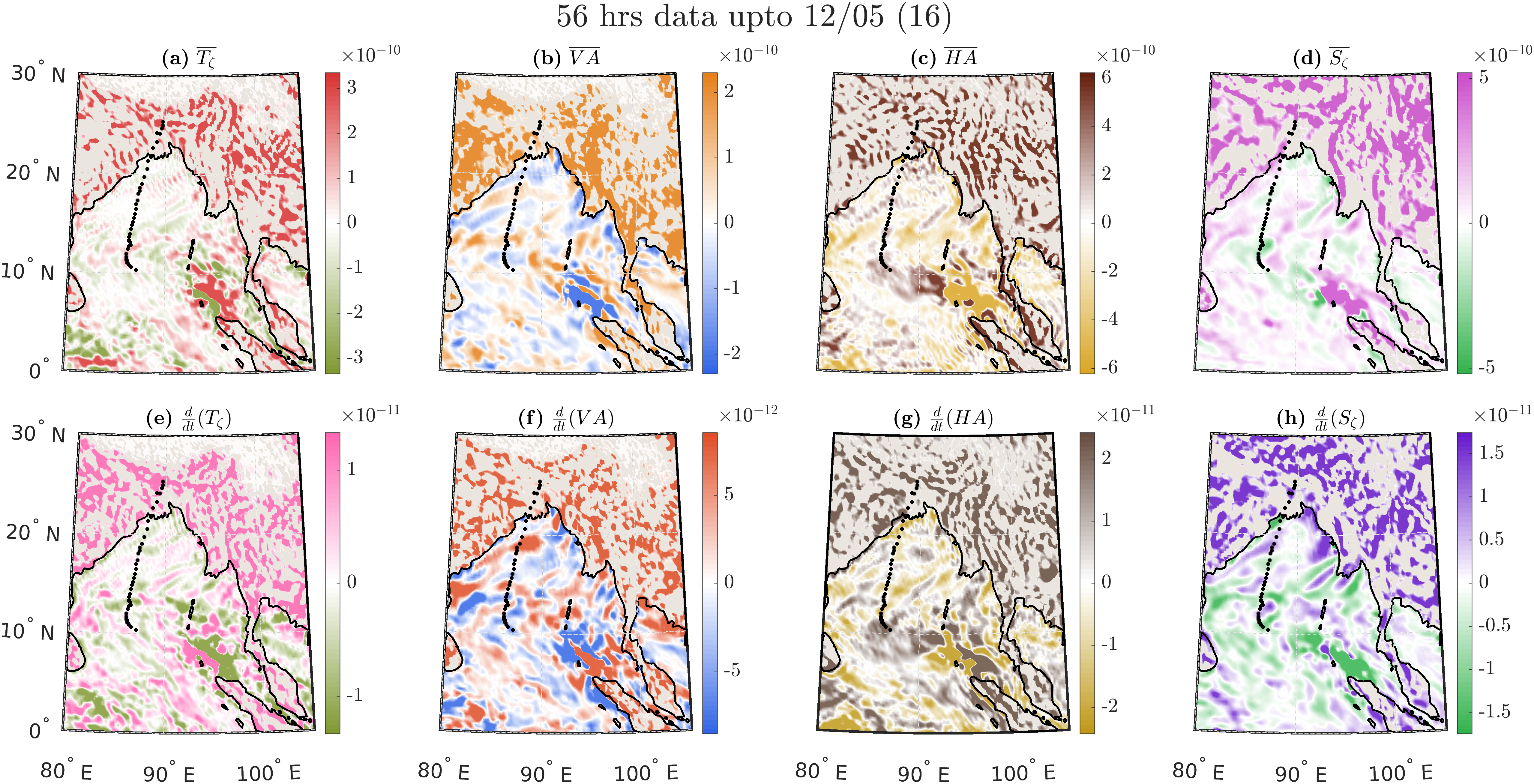}
    \caption{Vorticity budget analysis \textbf{much prior to the formation of a low pressure anomaly}, during the period corresponding to figure \ref{fig_BMA_Amphan_genesis}(a). Spatial distribution of $56$-hour mean of (a)tilting term $\overline{T_\zeta}$, (b) vertical advection $\overline{VA}$, (c) horizontal advection $\overline{HA}$, and (d) stretching term $\overline{S_\zeta}$. (e-f) The trend ($d/dt$) of terms in (a-d) respectively.}
    \label{fig_vortbudg_a_k4}
\end{figure}

\begin{figure}
    \centering
    \includegraphics[width=1\linewidth]{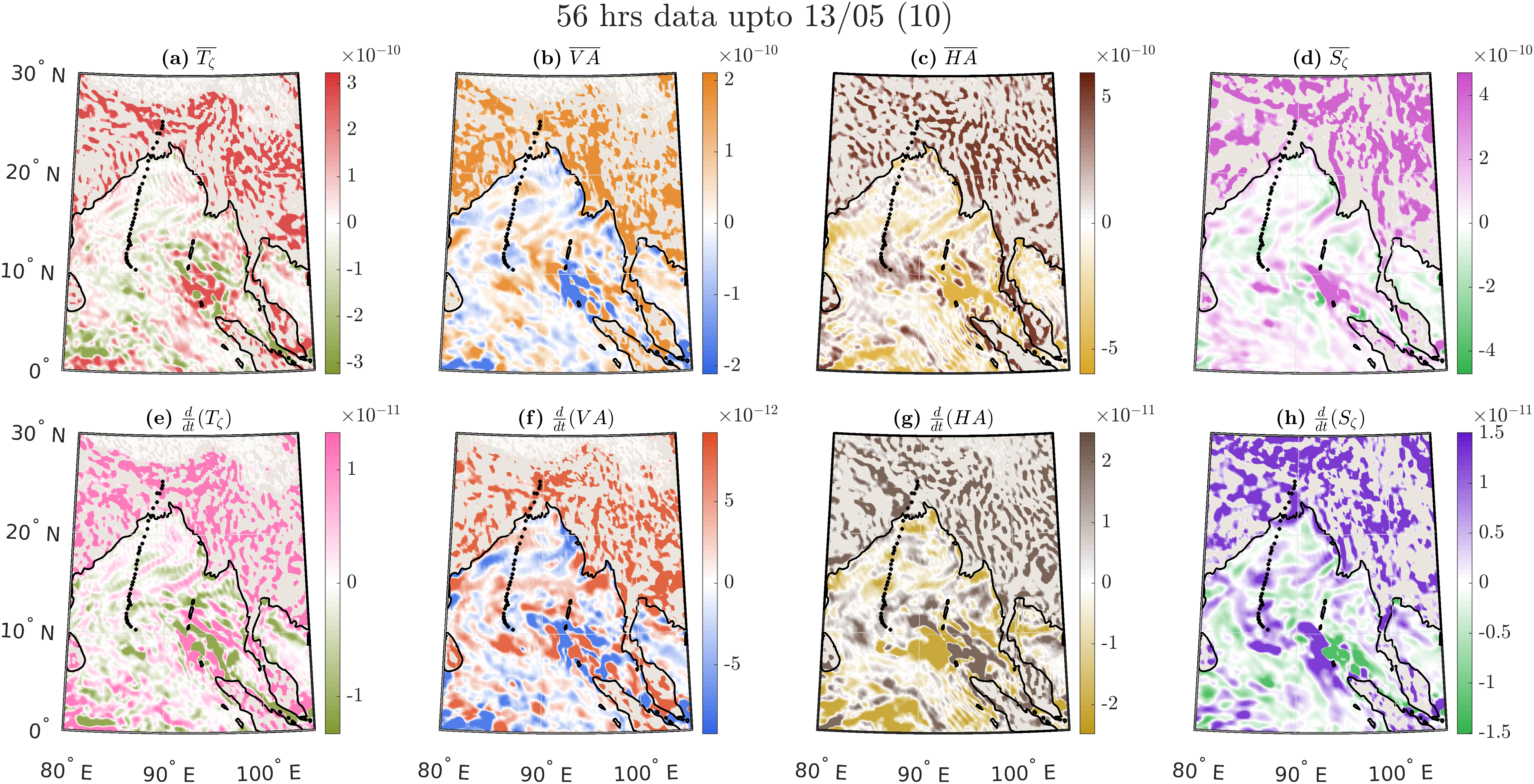}
    \caption{Vorticity budget analysis \textbf{much prior to the formation of a low pressure anomaly}, during the period corresponding to figure \ref{fig_BMA_Amphan_genesis}(b). Spatial distribution of $56$-hour mean of (a)tilting term $\overline{T_\zeta}$, (b) vertical advection $\overline{VA}$, (c) horizontal advection $\overline{HA}$, and (d) stretching term $\overline{S_\zeta}$. (e-f) The trend ($d/dt$) of terms in (a-d) respectively.}
    \label{fig_vortbudg_b_k10}
\end{figure}

\begin{figure}
    \centering
    \includegraphics[width=1\linewidth]{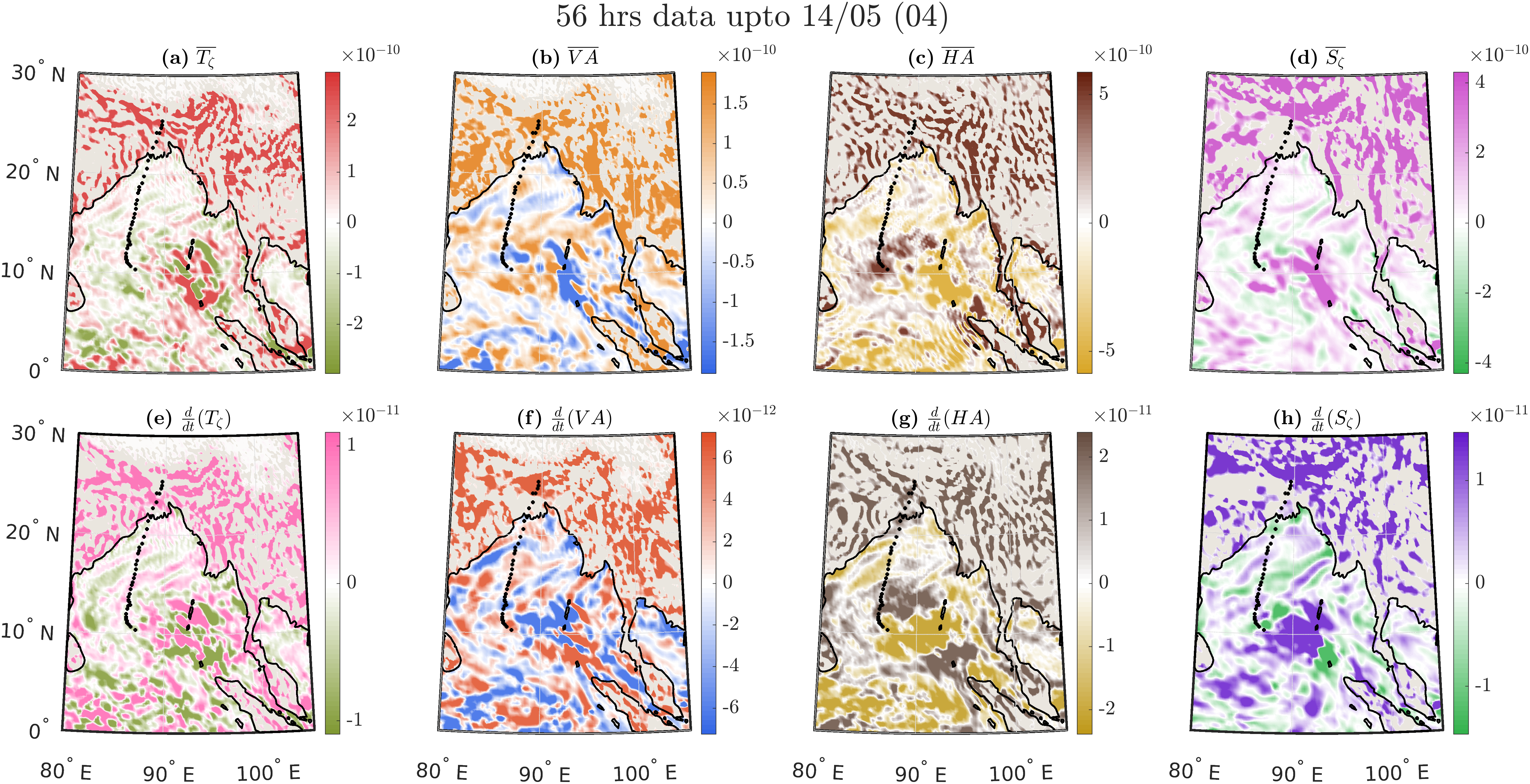}
    \caption{Vorticity budget analysis \textbf{during the formation of a low pressure anomaly}, during the period corresponding to figure \ref{fig_BMA_Amphan_genesis}(c). Spatial distribution of $56$-hour mean of (a)tilting term $\overline{T_\zeta}$, (b) vertical advection $\overline{VA}$, (c) horizontal advection $\overline{HA}$, and (d) stretching term $\overline{S_\zeta}$. (e-f) The trend ($d/dt$) of terms in (a-d) respectively.}
    \label{fig_vortbudg_c_k16}
\end{figure}

\begin{figure}
    \centering
    \includegraphics[width=1\linewidth]{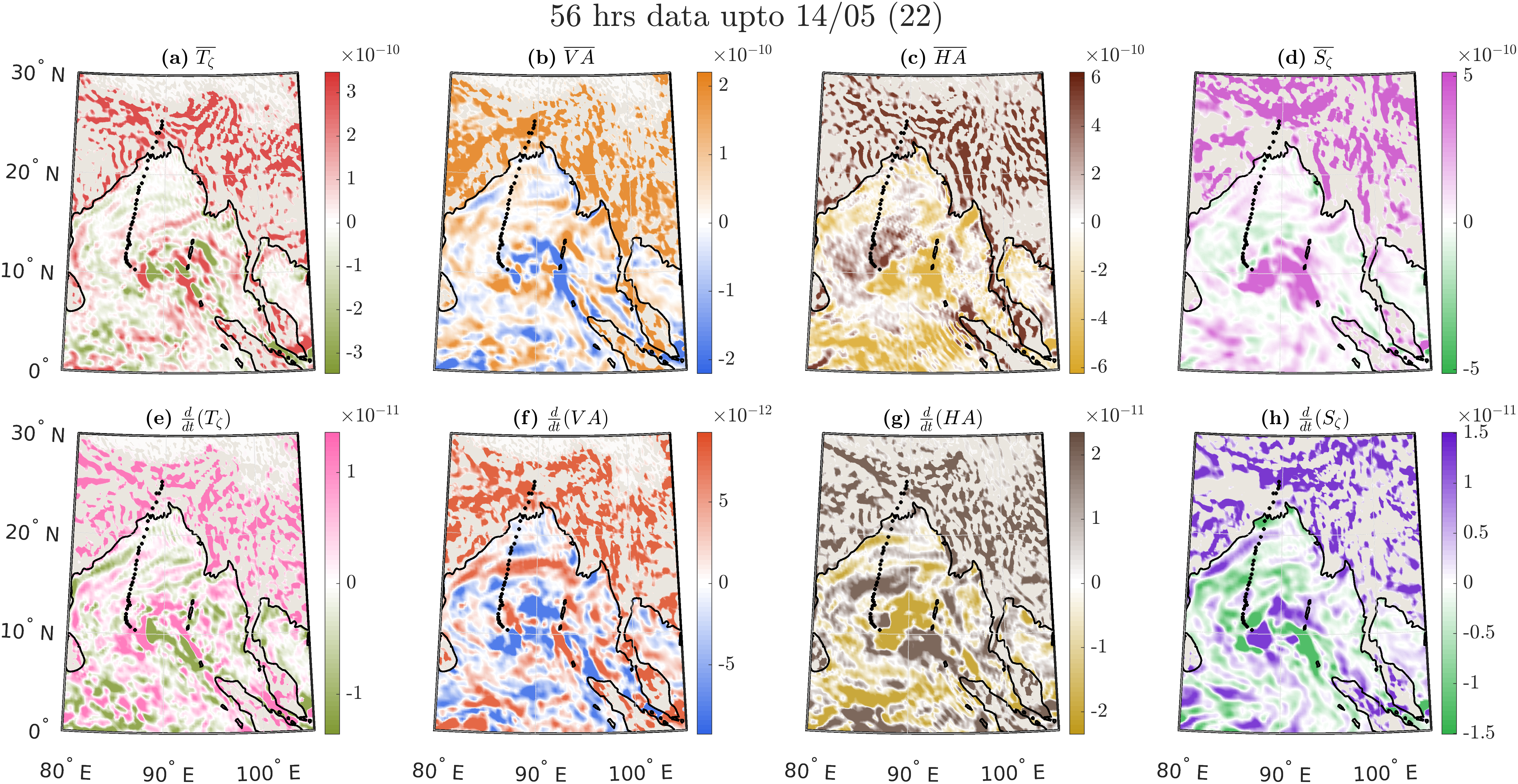}
    \caption{Vorticity budget analysis \textbf{during the formation of a low pressure anomaly}, during the period corresponding to figure \ref{fig_BMA_Amphan_genesis}(d). Spatial distribution of $56$-hour mean of (a)tilting term $\overline{T_\zeta}$, (b) vertical advection $\overline{VA}$, (c) horizontal advection $\overline{HA}$, and (d) stretching term $\overline{S_\zeta}$. (e-f) The trend ($d/dt$) of terms in (a-d) respectively.}
    \label{fig_vortbudg_d_k22}
\end{figure}

\begin{figure}
    \centering
    \includegraphics[width=1\linewidth]{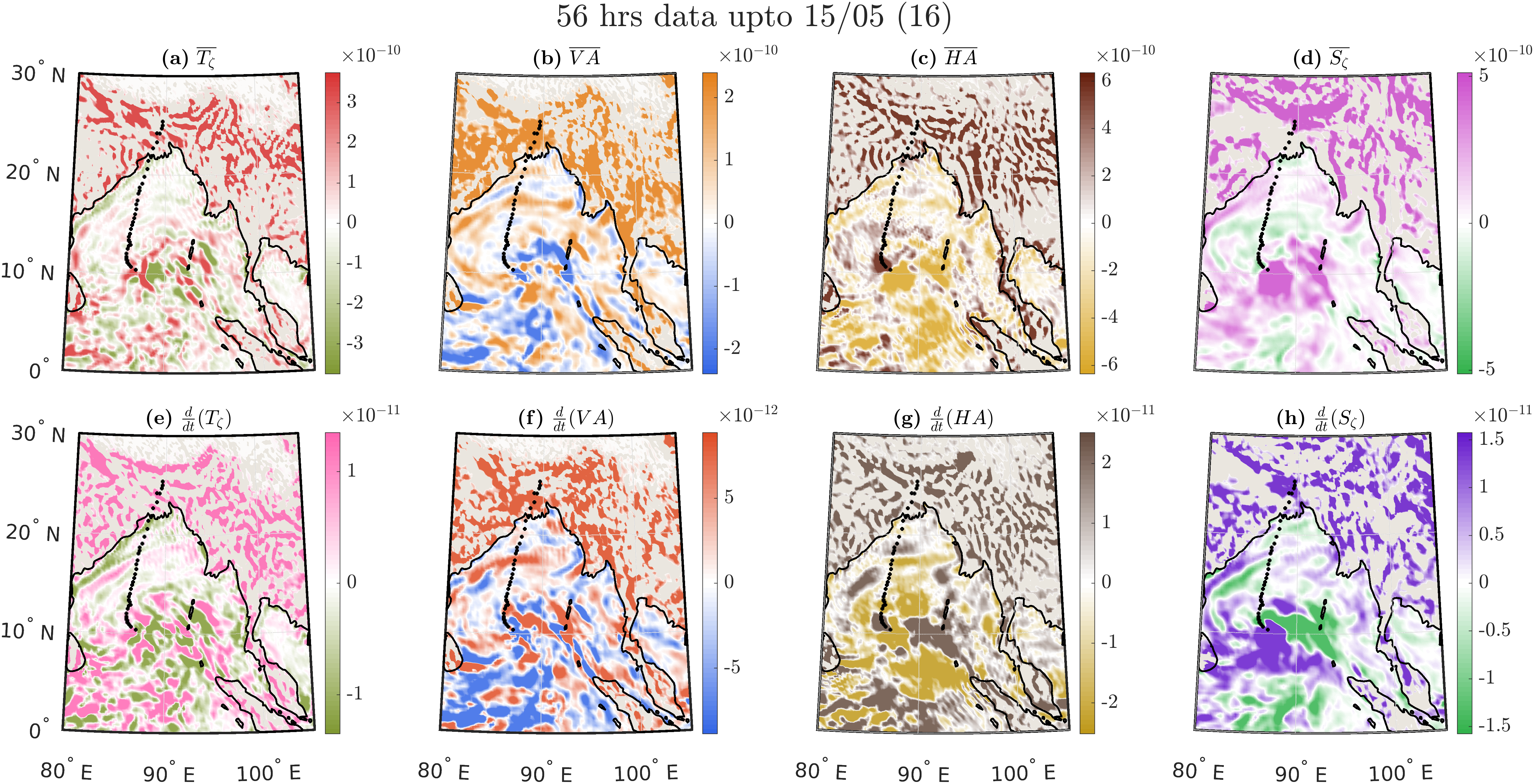}
    \caption{Vorticity budget analysis \textbf{during the intensification of the low pressure anomaly into a depression}, during the period corresponding to figure \ref{fig_BMA_Amphan_genesis}(e). Spatial distribution of $56$-hour mean of (a)tilting term $\overline{T_\zeta}$, (b) vertical advection $\overline{VA}$, (c) horizontal advection $\overline{HA}$, and (d) stretching term $\overline{S_\zeta}$. (e-f) The trend ($d/dt$) of terms in (a-d) respectively.}
    \label{fig_vortbudg_e_k28}
\end{figure}

\begin{figure}
    \centering
    \includegraphics[width=1\linewidth]{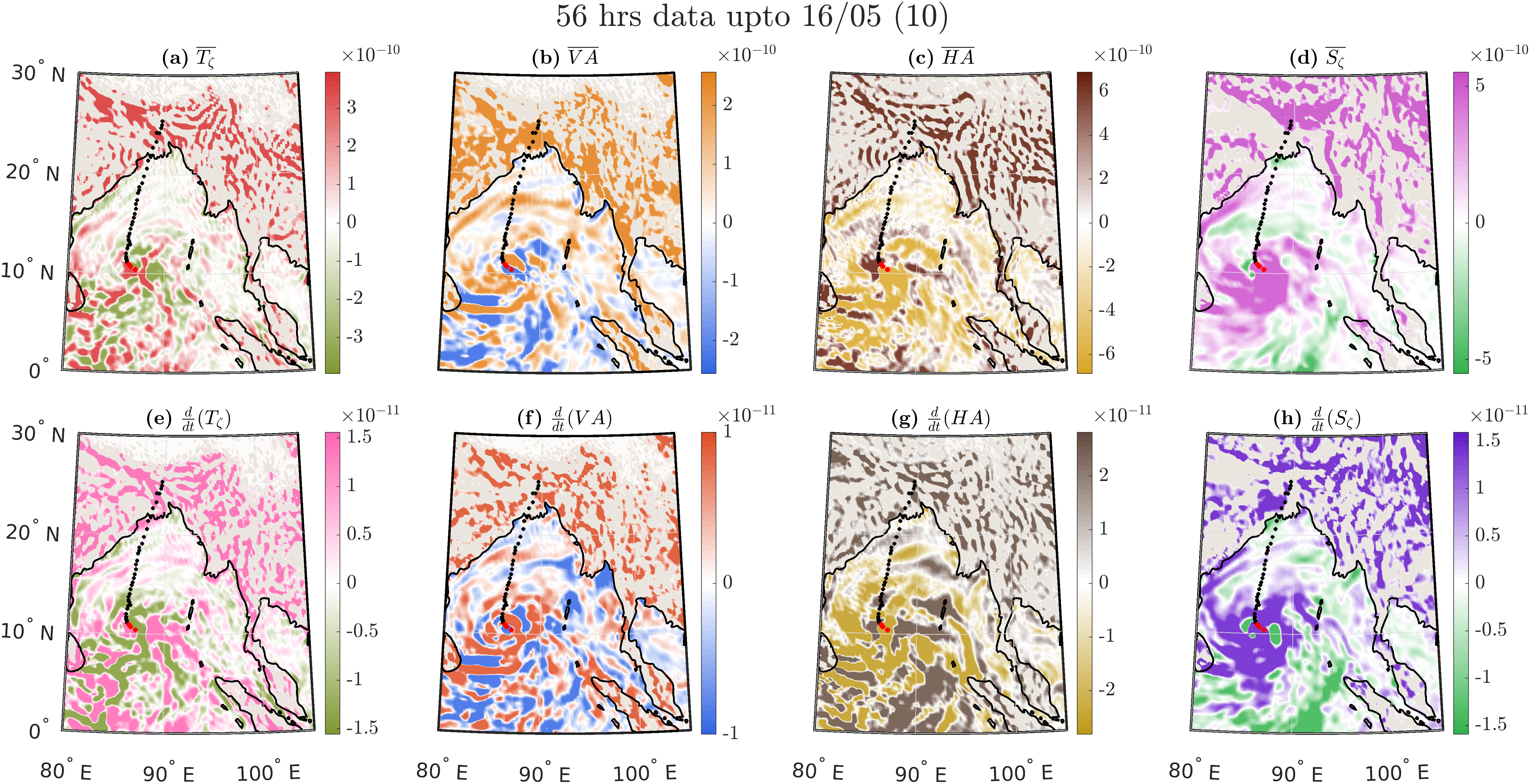}
    \caption{Vorticity budget analysis \textbf{during the intensification of the depression into cyclone Amphan}, during the period corresponding to figure \ref{fig_BMA_Amphan_genesis}(f). Spatial distribution of $56$-hour mean of (a)tilting term $\overline{T_\zeta}$, (b) vertical advection $\overline{VA}$, (c) horizontal advection $\overline{HA}$, and (d) stretching term $\overline{S_\zeta}$. (e-f) The trend ($d/dt$) of terms in (a-d) respectively.}
    \label{fig_vortbudg_f_k34}
\end{figure}

To understand how patches of high node strengths foster interdependence between regions of convection potential and inhibition during cyclogenesis, we perform a modal analysis of network connectivity evolution as described in section \ref{sec_methods}. This analysis reveals the nodes that are most influential (broadcast mode, $b$) and most influenced (receiving mode, $r$) during the time period corresponding to a set of ten consecutive networks (i.e., $56$ hours of data). Here, we examine the broadcast and receiving modes during specific $56$-hour periods prior to the formation of tropical cyclone Amphan in figure \ref{fig_BMA_Amphan_genesis}. Furthermore, we examine the contribution from different terms in the vorticity equation, Eq. \ref{eq_vortbudget_FINAL}, during these specific $56$-hour periods. Specifically, we plot the $56$-hour mean of the tilting term ($\overline{T_\zeta}$), the stretching term ($\overline{S_\zeta}$), and the horizontal and vertical advection terms ($\overline{HA}$ and $\overline{VA}$, respectively) in row-I of figure \ref{fig_vortbudg_a_k4}-\ref{fig_vortbudg_f_k34}. In row-II of figure \ref{fig_vortbudg_a_k4}-\ref{fig_vortbudg_f_k34}, we plot the linear trend (denoted by $d/dt$ of the respective variable) of each term in the vorticity budget equation.
We find that, much prior to the formation of a depression, the broadcast mode shown in figure \ref{fig_BMA_Amphan_genesis}(a) is significant in northern BoB in spatial pockets where CAPE is high but CI is weak, while the receiving mode occurs in neighboring patches where CAPE and CI are simultaneously high. Within an eighteen-hour period, the broadcast mode appears in northwestern BoB where CAPE is high, while the receiving mode expands from northern to central BoB in figure \ref{fig_BMA_Amphan_genesis}(b), where CI is prominent but has weakened compared to the previous period. Further, we observe that the regions where CI is weak (see figure \ref{fig_BMA_Amphan_genesis}(b)) are associated with multiple small disconnected patches of high node strength (see figure \ref{fig_NSvort_formn}(a,b). Since the broadcast and receiving modes appear to be significant in similar spatial pockets, while patches of high node strengths remain small and disconnected, we infer that the vorticity interactions remain localized during this period. The modal analysis reveals that patches of coherent vorticity dynamics occur initially in regions of high CI and high CAPE; however, such patches influence the vorticity in patches in neighboring locations where CI weakens. From vorticity budget analysis during this period, we find that patches of positive tilting and stretching occur in central BoB, and these patches are also associated with negative horizontal and vertical advection (see southeast BoB region in figure \ref{fig_vortbudg_a_k4} and \ref{fig_vortbudg_b_k10}). Although positive tilting and stretching can contribute to cyclonic vertical vorticity, negative $VA$ opposes the sustenance of cyclonic vorticity. A clear relation of the broadcast and receiving modes with the vorticity budget terms is therefore elusive during this period when a low pressure anomaly is not evident.


Subsequently, a low pressure anomaly forms in the southeast BoB. During this period, we observe from figure \ref{fig_BMA_Amphan_genesis}(c,d) that the broadcast and receiving modes are spatially spread-out across the entire basin. We also observe that locally coherent vorticity dynamics organize into large connected patches during this stage, as evident in figure \ref{fig_NSvort_formn}(c-e). 
The broadcast mode in figure \ref{fig_BMA_Amphan_genesis}(c) is relatively higher in northwest, central and southeast BoB, while the receiving mode is prominent in central and southeast BoB. We hypothesize that localized flow interactions are prominent in the south-east BoB, but also nonlocal influences may occur in a large spatial domain during this time period. The vorticity budget corresponding to figure \ref{fig_BMA_Amphan_genesis}(c) is shown in figure \ref{fig_vortbudg_c_k16}. We find multiple patches of both positive and negative $\overline{VA}$, patches of positive stretching, and a positive trend of vertical advection in the central and southeast BoB (figure \ref{fig_vortbudg_c_k16}) where both broadcast and receiving modes are prominent. 
In the northwest and central BoB region, we find that the mean and trend of vertical advection are positive as well as patches of positive stretching and negative $\overline{HA}$ appear.

In figure \ref{fig_BMA_Amphan_genesis}(d), we find that the broadcast mode becomes prominent in the southeast BoB region, similar to where the receiving mode occurred in the previous period in figure \ref{fig_BMA_Amphan_genesis}(c). On the other hand, the receiving mode develops in northwest BoB in regions where CI is high but weakens over the subsequent periods. Clearly, the most influenced and influential regions are spatially separated, indicating nonlocal effects emerging from local vorticity interactions. During this period, the receiving mode is associated with positive means and positive trends of both $VA$ as well as $HA$ in northwest and central BoB, while tilting and stretching are weak in this region (figure \ref{fig_vortbudg_d_k22}). Further, the broadcast mode is associated with multiple small patches with positive mean and trend of tilting, increasing tendency for $VA$ (positive $d{VA}/dt$) and negative $\overline{HA}$.

As the low pressure system attains the strength of a depression, we observe that the broadcast mode is spatially spread throughout the basin, but the receiving mode is concentrated in southwest BoB region as evident in figure \ref{fig_BMA_Amphan_genesis}(e). Vorticity budget analysis (see figure \ref{fig_vortbudg_e_k28} reveals that the receiving mode occurs in regions where titling is negligible or occurs in incoherent patches, but the tendency for tilting to increase is high (positive trend). Instead, patches of positive tilting appear to organize in another region in central BoB in a neighboring region east of where the receiving mode occurs. Also, stretching is positive, and the tendency for both stretching and $VA$ to increase is high in southwest BoB where the receiving mode occurs. Interestingly, here the receiving mode occurs in a region where CI has already weakened during previous periods. However, we notice that the contours of CAPE in the region of the receiving mode shift from relatively low values in figure \ref{fig_BMA_Amphan_genesis}(e), to higher values appear in the subsequent period (compare the contours of CAPE in southwest BoB across figure \ref{fig_BMA_Amphan_genesis}(e-f)).

Finally, as the low pressure system forms a depression and intensifies into a cyclone (refer figure \ref{fig_BMA_Amphan_genesis}(f)), we observe that the broadcast mode occurs in the central and southwest BoB region, similar to the receiving mode of the previous period. 
The broadcast mode occurs in regions devoid of convective inhibition and with high CAPE while the receiving mode occurs in eastern BoB along the coasts where CI is weakening. We also observe that the broadcast mode is associated with positive means and trends of tilting, $VA$, and stretching terms, but negative mean of $HA$ (see figure \ref{fig_vortbudg_f_k34}). As the southwest BoB region converts from a receiving mode to a broadcast mode, we observe that more patches of positive $\overline{VA}$, negative $\overline{HA}$, positive $\overline{S_\zeta}$ and positive $\overline{T_\zeta}$ appear in this region (compare figures \ref{fig_vortbudg_e_k28} and \ref{fig_vortbudg_f_k34}).
Although, some coherent pockets of positive mean and trend of $HA$ also occur in the central BoB region indicating entrainment of neighboring moist wind by the cyclone. Stretching increases significantly and occurs in a large coherent pocket in central and western BoB in figure \ref{fig_vortbudg_f_k34} as compared to that in figure \ref{fig_vortbudg_e_k28}.

In summary, vorticity interactions are localized during initial stages in the absence of a low pressure anomaly. During the intensification of a low pressure anomaly into a depression and subsequently into a cyclone, we make the following general observations: (i) The broadcast mode occurs most likely in regions where high CAPE exists, is most likely associated with regions of highly negative $\overline{HA}$, positive $\overline{T_\zeta}$ and $d\overline{T_\zeta}/dt$, positive $\overline{VA}$. A positive trend of stretching is also likely to occur in regions highlighted by the broadcast mode. (ii) The receiving mode is likely to occur in regions where either CI is weakening or CAPE is increasing, and is associated with positive $\overline{VA}$, positive $d{VA}/dt$, but negligible tilting. The receiving mode may also be associated with negative $\overline{HA}$ during some periods. (iii) Another important observation is that the regions highlighted by the receiving mode of a specific period tend to become the broadcast mode of the subsequent period. These observations help us infer the physical mechanism as explained in the following subsection. 

\subsection{A general understanding of large-scale emergence and modes of small-scale interactions}\label{sec_results_general}

Intriguingly, regions of high CAPE and high CI co-evolve in an interdependent manner during cyclogenesis (evident via figure \ref{fig_TS_ntwk}(d,e)). This co-evolution is facilitated by patterns of locally coherent vorticity dynamics that evolve through large-scale modes (shown in figure \ref{fig_BMA_Amphan_genesis}). To understand the generating mechanism that produces such large-scale patterns, consider the vorticity field set up in the turbulent background flow by tropical perturbations such as an easterly wave. In such flows, vorticity dynamics can remain correlated for short periods due to local interactions in regions of strong shear or perturbations \citep{MartinJFM2021synchronisation}. Initially, we find small patches of coherent vorticity dynamics appear and disappear randomly at diverse locations. 
Moisture convergence cannot be sustained in regions of high CI. However, if CAPE is high in a wide neighborhood and a patch of coherent vorticity dynamics occurs, then short-lived convergence can be induced and moist air will be entrained. Such accumulation of moisture can enhance the moisture feedback in adjacent locations, which further enhances vorticity anomalies and promotes convergence locally \citep{wing2018convective}. Therefore, coherence of vorticity can be sustained locally and can expand to adjacent locations, even in the absence of a large-scale low pressure anomaly.

Much prior to the formation of a low pressure anomaly, vorticity interactions are localized and uncoordinated across the domain. In some spatial pockets, CAPE builds up locally but convective outbursts are suppressed by high convective inhibition. Such regions experience updrafts, but do not sustain low-level convergence. Whether low-level convergence amplifies cyclonic vorticity is measured by the stretching term. We find such regions of high CI and strong $VA$ are also associated with low stretching and tilting, and generally comprise the receiving mode. We observe that eventually, convective inhibition weakens when the updrafts are able to sustain low-level convergence and vertical advection of cyclonic moist wind. This however requires sustained generation of updrafts in regions of high CI. How are strong updrafts sustained in the regions of the receiving mode, can be explained using nonlocal interactions across regions of receiving and broadcast modes, as follows. 

Nonlocal effects occur in turbulent velocity field, as also discussed by \citet{mukherjee2022correlation}. We note that tilting, stretching or vertical advection represent \textit{local} evolution of local vertical vorticity and cannot explain how nonlocal influences appear between regions that are spatially separated. 
Interestingly, this can be explained using the horizontal advection term. Positive $\overline{HA}$ characterizes regions where cyclonic vorticity is drawn inwards towards a central core contributing to cyclonic spin. We observe positive $\overline{HA}$ in small coherent pockets when the cyclone is formed (see figure \ref{fig_vortbudg_e_k28}(c) and \ref{fig_vortbudg_f_k34}(c)), highlighting regions where moist updrafts are drawn into the cyclone core. On the contrary, negative $\overline{HA}$ highlights regions where cyclonic vertical vorticity is pushed away from the core. We notice that $\overline{HA}$ is generally negative in regions that are most influential (broadcast mode) during a selected period. We can thus infer that the broadcast mode highlights regions where not only is the vertical vorticity amplified by stretching, tilting and vertical advection, but also the region from where such strong vertical vorticity advects to other locations. 

Advection of cyclonic vertical vorticity from regions of high CAPE (broadcast mode) to regions of high CAPE and CI (receiving mode) help initiate and sustain strong updrafts in regions of the receiving mode and weaken the local CI or increase the local CAPE. Feedback between moist convection and vorticity coherence further weakens CI locally; weakening of CI is evident from the reduced spatial extent of CI across the subplots in figure \ref{fig_capeci_all}. As very strong convective outbursts occur, the strong updrafts interact with horizontal vortical elements and tilt their direction to contribute to vertical vorticity. As a result, positive tilting appears along with very strong updrafts in regions where CI has weakened significantly and CAPE is high. Similarly, once CI weakens, low-level convergence of moist wind is sustained due to high CAPE, and reflects as increase in the stretching term. These characteristics--- positive $VA$, tilting and stretching--- are attribute of regions highlighted by the broadcast mode. Therefore, a region which was initially the receiving mode, translates into the broadcast mode of the subsequent period. Again, owing to negative $HA$, these regions influence other regions and produce new receiving modes. In fact, we can infer that the nonlocal influence and large-scale patterns emerge when outward (negative) horizontal advection advects vertical vorticity across regions, and aids the occurrence of simultaneous self-organized convective outbursts in a large spatial domain.

This also explains why we observe an increase in $f_{CI,NS}$ and $\langle CI \rangle _{HNS}/\langle CI\rangle$ in figure \ref{fig_TS_ntwk}, just before the depression intensifies into a cyclone. As locally coherent vertical vorticity dynamics is transported from regions of high CAPE to regions with high CAPE and high CI, more patches of high node strengths become associated with high CAPE and high CI simultaneously. Hence, both $f_{CI,NS}$ and $f_{CAPE,NS}$ increase and become correlated. Evidently, the local coupling of moisture thermodynamic parameters (CAPE and CI) with local vorticity dynamics becomes spatially correlated over a large basin through the horizontal advection of vertical vorticity. Interestingly, our results in figure \ref{fig_TS_ntwk} show that such spatially coordinated moisture convection remains sustained only when a depression forms a cyclone, and not otherwise. In summary, we explain the physical mechanism of how large-scale patterns emerge from the self-organization of local moisture convection and vortical interactions, that differentiate developing and non-developing storms.

\section{Conclusion}

Under the purview of complex systems theory, we view the formation of a cyclone as the emergence of a large-scale self-organized pattern of moist convection. Our examination of tropical cyclogenesis in Bay of Bengal reveals novel multi-scale interdependence between the dynamical organization of local vortical interactions and moist air thermodynamics. Intriguingly, a large-scale pattern emerges \textit{spontaneously} from the arrangement of several patches of locally coherent vorticity dynamics. Moisture convection depends on the local enhancement of vertical vorticity which also depends on local vorticity interactions. Thus we examined the thermodynamic parameters of moisture convection in regions of high node strengths. 

We find that the fractional intensity of CAPE associated with regions of strong local vorticity interactions increases before a cyclone forms. Counterintuitively, the fractional intensity of convective inhibition (CI) in regions of coherent vorticity dynamics also increase prior to cyclogenesis. Furthermore, the spatial spread (number) of locations where vorticity interactions and CAPE (or CI) are simultaneously high, increase prior to cyclogenesis. We discover an intriguing spatial correlation consistent in multiple cases where depressions form cyclones--- the patches of coherent vorticity dynamics associated with high CAPE evolve almost synchronously with patches of coherent vorticity dynamics associated with high CI, just prior to and during cyclogenesis. Such correlations are negligible for depressions that do not form cyclones. Even if such spatial correlation arises for a non-developing depression initially, the correlation declines rapidly in the subsequent period. This finding suggests that an interdependence between diverse locations of convection potential and inhibition arises through large-scale `communication' and organization in the flow. To understand such cross-regional interdependence in local moist thermodynamic processes, we perform broadcast mode analysis and examine the contribution of different physical mechanisms that enhance vertical vorticity.

Modal analysis of network evolution reveals that the local coherence in vorticity dynamics propagates between different regions across the flow domain. Propagation of locally coherent vorticity dynamics is facilitated by large-scale modes, consisting of most influential (broadcast mode) and most influenced (receiving mode) regions of the flow. Analyzing the typical characteristics of the vorticity field in locations of the broadcast and receiving mode, we are able to explain the physical process of self-organization of local moist convection over a large spatial domain. Outward horizontal advection of vertical vorticity from the most influential regions influence and help sustain strong vertical vorticity in most influenced regions, thus reducing convective inhibition in these regions. Vertical vorticity is continually enhanced by local tilting, stretching and vertical advection in regions of high CAPE and where CI weakens.
Thus, both local amplification of vertical vorticity and nonlocal influence occur through horizontal advection in the velocity field. Local thermo-fluid interactions that occur between coherent vorticity patches and moisture aggregation, also become organized at large scales. Such spatial correlations are sustained only when a depression develops into a cyclone and not otherwise. 

Further, we can understand both \textit{emergence} and \textit{self-organization} during cyclogenesis. The phenomenon of \textit{emergence} is evident from the synoptic patterns formed from flow interactions within sub-meso scales. Insights from complexity science dictate that during emergence, a two-way interdependence is created between the large and small scales in the system; the emergent pattern influences the processes at smaller scales \citep{goldstein1999emergence}. 
During cyclogenesis, local vorticity interactions at the small-scale organize into patterns that in turn facilitate the non-local communication of local vorticity interactions, that in turn influence moisture convergence at smaller scales. 
\textit{Self-organization} occurs when interactions in a complex system can autonomously increase order and foster a novel process in the system \citep{prigogine1985self, haken2006information}. During cyclogenesis, self-organization is evident since local vorticity interactions and convective outbursts are spatially correlated across a large domain; this delineated novel multi-scale therm-fluid interactions during cyclogenesis. The physical mechanism of self-organization is understood through modal analysis and vorticity budgets presented in this work. Further, self-organization induces a crucial interdependence between the thermodynamic parameters of moist convection and vorticity dynamics, that can distinguish developing and non-developing storms.

A promising direction for future is to use the understanding developed in this work to formulate predictive measures that can distinguish developing and non-developing cases. The effect of seasonality on the relation between vorticity dynamics and convection potential and inhibition can be investigated in future by analyzing cyclones that form during winter months in the Bay of Bengal. Further, using such analysis to compare cyclogenesis in different oceanic basins can shed light on role of background vorticity in determining the intensity of cyclones. Another interesting realm for future works is to examine cyclogenesis by modeling the multi-scale interdependence between vorticity dynamics and moist convection. 

\vspace{10pt}

\textbf{Acknowledgements}{This work is funded from the IoE initiative (SP/22-23/1222/CPETWOCTSH). S.T. acknowledges the support from Prime Minister Research Fellowship, Govt. of India. B.N.G. acknowledges the support from ANRF Prime Minister Professorship.}

\textbf{Supporting Information} {Includes two Supplementary Movies.} 

\textbf{Author Contributions} {S.T.: conception(lead), writing (lead), analysis and visualization (lead), interpretation (lead); A.S.: analysis (equal), editing (equal); B.N.G.: conception(equal), editing (equal), interpretation (equal); R.I.S.: conception (equal), editing (lead), interpretation (equal), funding (lead).}

\textbf{Competing Interests} {The authors declare that they have no conflict of interests.}

\textbf{Data and materials availability} {All the information necessary for evaluating the results of the manuscript are available in the main text and supplementary. The source of data of climate variables is appropriately cited in the Methods section and a code for network construction with GPU acceleration is made available on \\ https://github.com/shrutitandon2111/BMAcyclone.git.}

\appendix

\section{Effect of spatial proximity constraint in studying the emergence of large-scale patterns }\label{SIsec_boxsize}

In this work, we have investigated the emergence of large-scale patterns while encoding small-scale vorticity interactions between geographical locations (nodes) that are in spatial proximity. In the main text, we have shown results for the case when such interactions are encoded between nodes that are separated by a maximum distance of $\approx 220$ km ($\leq 2^\circ$ in latitude or longitude). Here, we show that small changes in the spatial proximity constraint do not change the qualitative nature of the emergent large-scale patterns.

\begin{figure}[h!]
    \centering
    \includegraphics[width=1\linewidth]{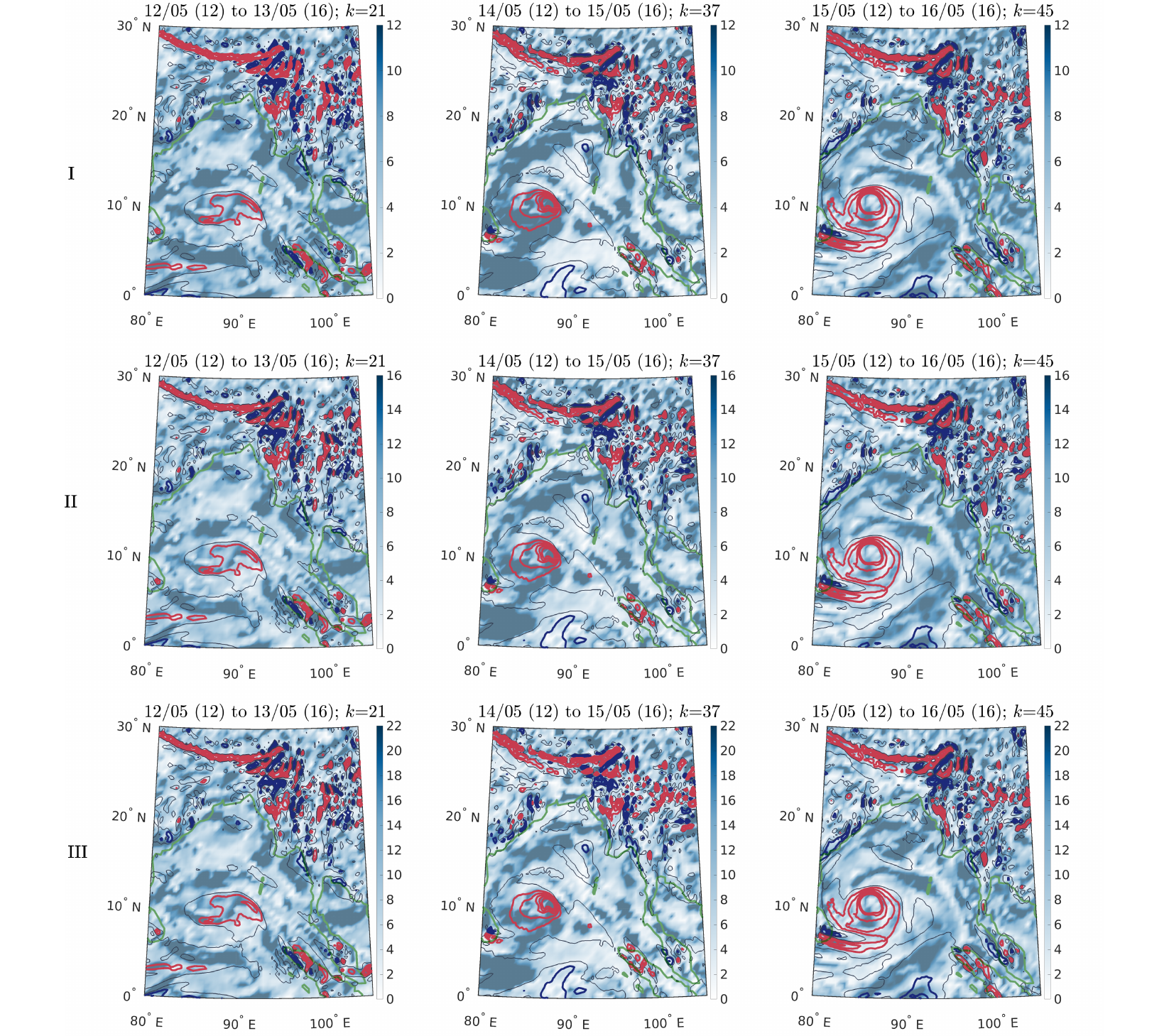}
    \caption{Spatial node strength distribution obtained during occurrence of cyclone Amphan. Nodes are connected within a maximum separation of $\leq 1.5^\circ$ (in row-I), $\leq 1.75^\circ$ (in row-II), and $\leq 2^\circ$ (in row-III) in latitude or longitude. The corresponding length scales of the spatial proximity constraint are $\approx 165$ km, $\approx 195$ km, and $\approx 220$ km, respectively. The patterns are identical qualitatively for different spatial proximity constraints.}
    \label{figsupp_spatprox}
\end{figure}

Figure \ref{figsupp_spatprox} shows the spatial distribution of node strengths obtained during the occurrence of cyclone Amphan for network time frames $k=21,~37,~45$ for different spatial proximity constraints. We have considered networks by connecting nodes within a maximum separation of (i) $\leq 1.5^\circ$, (ii) $\leq 1.75^\circ$, and (iii) $\leq 2^\circ$ in latitude or longitude, as shown in rows I, II, and III of figure \ref{figsupp_spatprox}, respectively. The large-scale patterns emerging from small-scale vorticity interactions are similar during each network time-frame despite changes in the spatial proximity constraint. We find that the range of the values of the node strengths varies owing to the changes in the number of nodes considered as neighbors by the modified spatial proximity constraint.

\section{Effect of spatial resolution of data on emergence of large-scale patterns }\label{SIsec_resolution}

In the main text, we construct networks from relative vorticity data at a spatial resolution of $0.5^\circ \times 0.5^\circ$ in latitude and longitude. Here, we examine the patterns that emerge in the network if we use a spatial resolution of $0.25^\circ \times 0.25^\circ$. The total number of nodes (including the extended domain in which the network is constructed as shown in Methods section) is $5751$ for $0.5^\circ \times 0.5^\circ$ resolution and $22701$ for $0.25^\circ \times 0.25^\circ$ resolution. Note that, due to limitations of computational memory and time, we cannot perform statistical significance tests for network connectivity at this resolution. Hence, we have analyzed the patterns obtained from $0.5^\circ \times 0.5^\circ$ in detail in the main text. We use the same spatial proximity constraint (neighbors $\leq 2^\circ$ in latitude or longitude) when constructing networks from both data resolutions.
\begin{figure}[h!]
    \centering
    \includegraphics[width=1\linewidth]{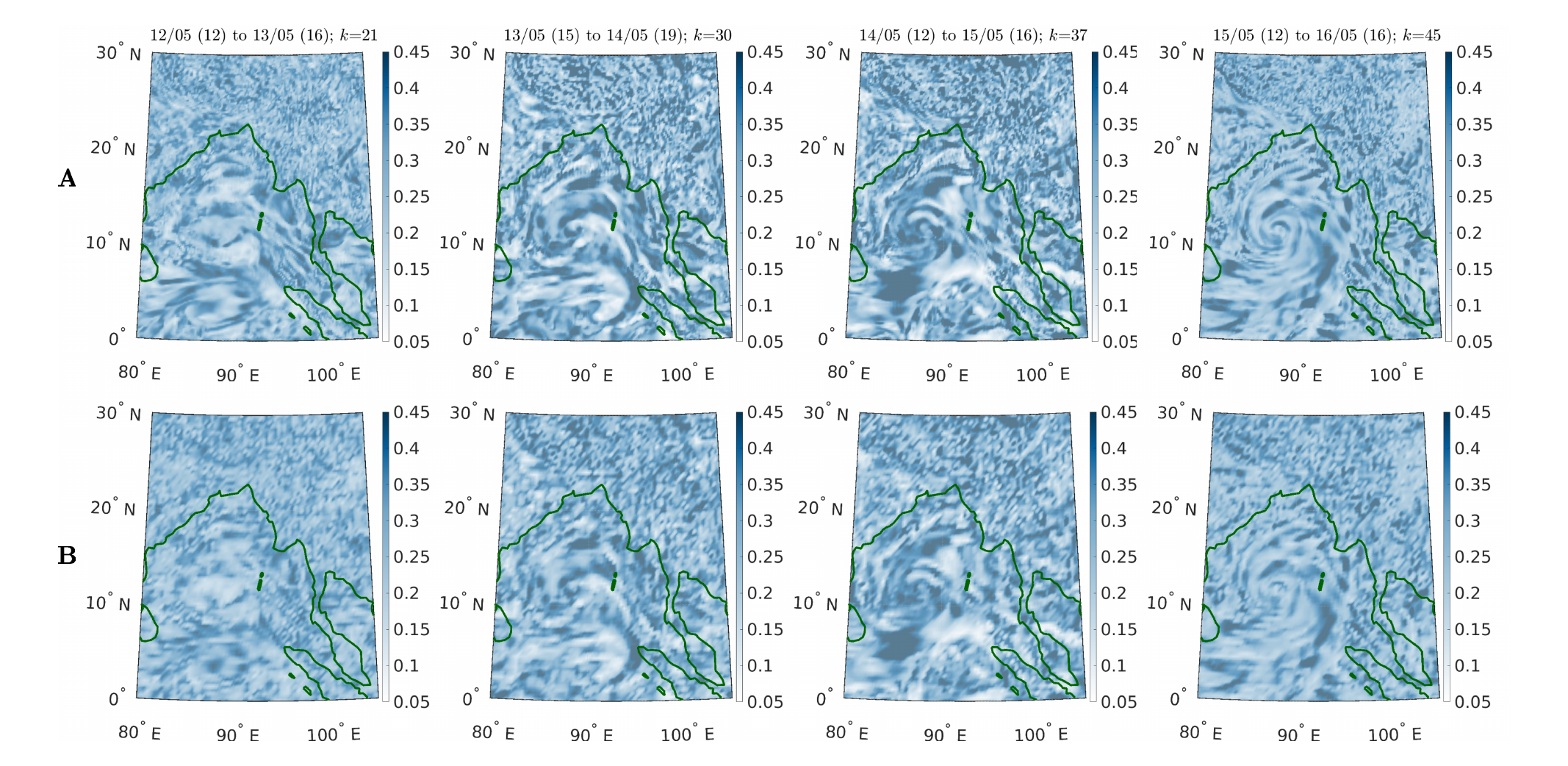}
    \caption{Spatial node strength distribution obtained from spatial-proximity networks for $k=21,~30,~37,~45$ of relative vorticity ($\omega$) when nodes are connected within a maximum separation of $2^\circ$ in latitude or longitude. The data is considered at resolution $0.25^\circ \times 0.25^\circ$ shown in row (A), and $0.5^\circ \times 0.5^\circ$ shown in row (B). For both resolutions, we find that large-scale patterns emerge from small-scale vorticity interactions. To facilitate comparison, the node strengths are normalized by the maximum number of neighbors in both cases. The maximum number of neighbors is essentially the maximum number of links a node can make based on the spatial proximity constraint and data resolution, which is $80$ for lower resolution (row A), and $288$ in case of higher resolution data (row B). Results are shown for networks where links are established without statistical significance for both resolutions.}
    \label{figsupp_pt25res}
\end{figure}

We find that for higher ($0.25^\circ \times 0.25^\circ$) resolution of data (see figure \ref{figsupp_pt25res} row-A), large-scale patterns are essentially identical to the patterns obtained from $0.5^\circ \times 0.5^\circ$ resolution, except that the patterns contain much more intricacies. For example, see the spatial distribution of node strengths in rows A and B for network $k=30$ and $k=37$ in figure \ref{figsupp_pt25res}. Large extended structures are evident for both data resolutions in the exact same locations. However, due to a higher resolution, the spiral arrangement of such extended structures is more prominent in figure \ref{figsupp_pt25res} row-A. Further, for $k=21$ and $k=45$, notice the patches of high node strengths that become prominent in the central Bay of Bengal region in figure \ref{figsupp_pt25res} row A, while it is not prominent in figure \ref{figsupp_pt25res} row B. Thus, at higher resolution, more intricate patterns emerge; computing the correlations of CAPE and CI associated with regions of high node strengths from networks constructed at higher resolutions can help further improve the correlation statistics and distinguish developing and non-developing cases better, as presented in the results section in the main text. Another interesting aspect for future work would be to relate the patterns emerging from turbulent flow interactions at much smaller scales such as the cumulus scales using cloud-resolving models.

\section{Spatial averages of network connectivity, CAPE, CI during developing and non-developing cases}\label{SIsec_capeciTS}

\begin{figure}[h!]
    \centering
    \includegraphics[width=1\linewidth]{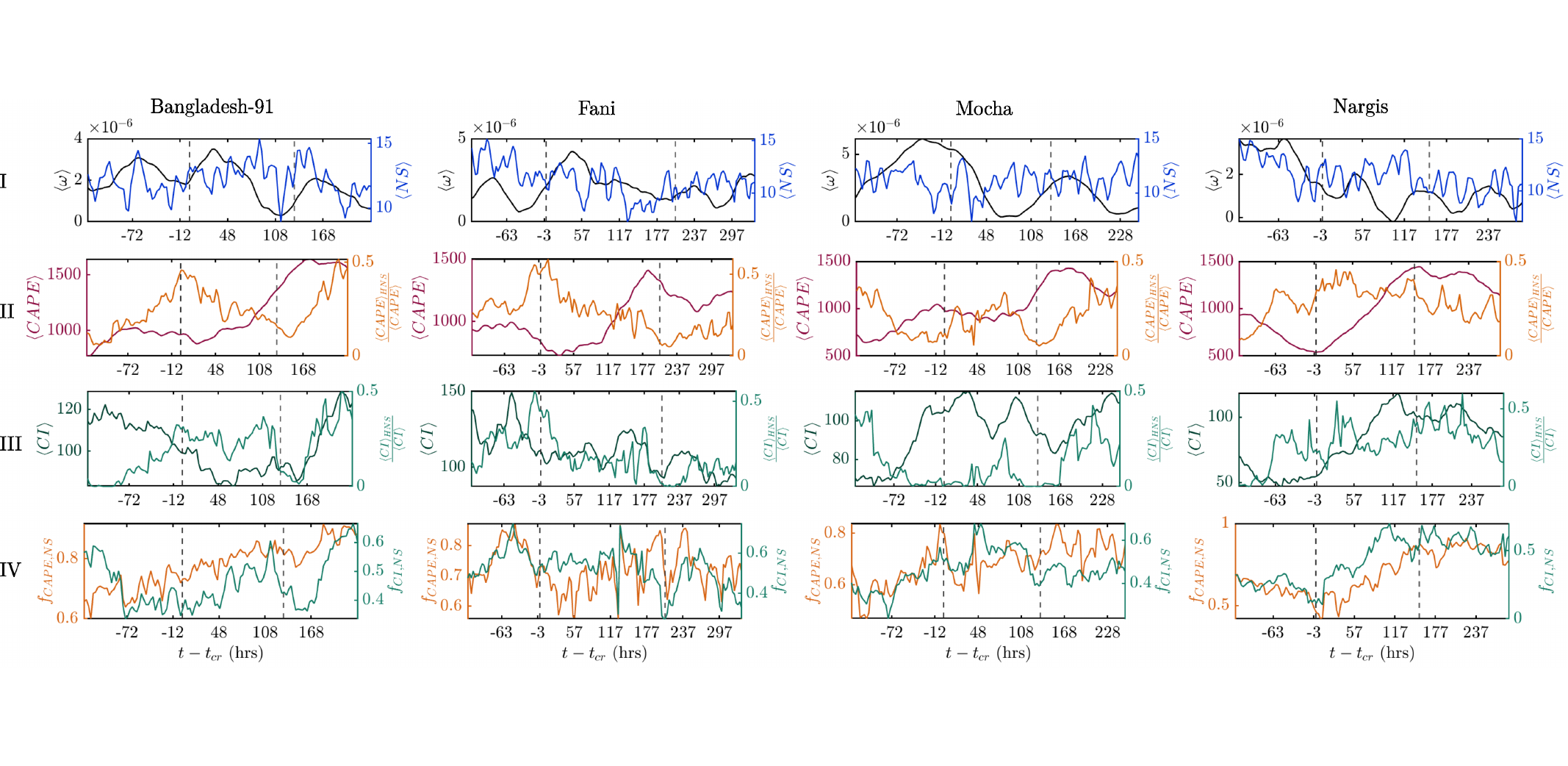}
    \caption{\textbf{Developing cases:} Row-(I): Temporal variation of the spatial average of vorticity strength ($\langle\omega\rangle$) and node strength ($\langle NS \rangle$) during each network time-frame. Row-(II): Overall spatially averaged intensity of convection available potential energy ($\langle CAPE\rangle$), and fractional intensity of convection potential ( $\langle CAPE \rangle _{HNS}/\langle CAPE\rangle$). Row-(III):  Evolution of overall spatially averaged intensity of convection inhibition ($\langle CI\rangle$), and fractional intensity of convection inhibition ( $\langle CI \rangle _{HNS}/\langle CI\rangle$), $NS>15$. Row-(IV): Variations of the number density of locations, $f_{CAPE,NS}$ and $f_{CI,NS}$ given thresholds $NS>15$, $CAPE>900$ J/kg,  and $CI>200$ J/kg. Subplots in each row are shown for various developing cases, and the names of the cyclones are mentioned above  the corresponding columns. The abscissa demarcates time relative to the critical time $t_{cr}$ (also indicated by a dashed gray line at zero), that is the time at which a depression is identified. The second dashed line indicates the time when the cyclone intensity began to weaken and dissipate.}
    \label{figsupp_dev_capeciTS}
\end{figure}

\begin{figure}
    \centering
    \includegraphics[width=1\linewidth]{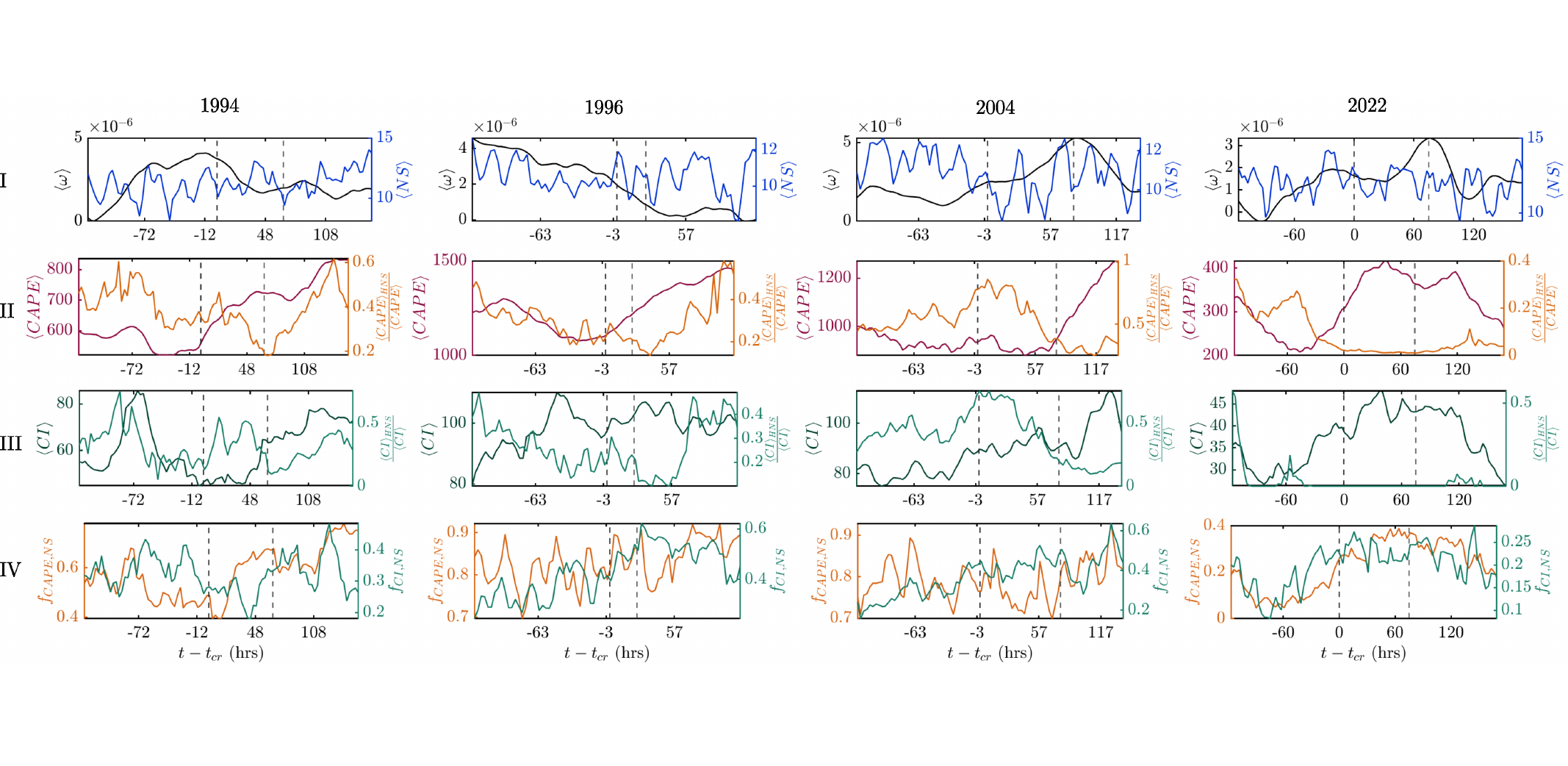}
    \caption{\textbf{Non-developing cases:} Row-(I): Temporal variation of the spatial average of vorticity strength ($\langle\omega\rangle$) and node strength ($\langle NS \rangle$) during each network time-frame. Row-(II): Overall spatially averaged intensity of convection available potential energy ($\langle CAPE\rangle$), and fractional intensity of convection potential ( $\langle CAPE \rangle _{HNS}/\langle CAPE\rangle$). Row-(III):  Evolution of overall spatially averaged intensity of convection inhibition ($\langle CI\rangle$), and fractional intensity of convection inhibition ( $\langle CI \rangle _{HNS}/\langle CI\rangle$), $NS>15$. Row-(IV): Variations of the number density of locations, $f_{CAPE,NS}$ and $f_{CI,NS}$ given thresholds $NS>15$, $CAPE>900$ J/kg,  and $CI>200$ J/kg. Subplots in each row are shown for various non-developing cases, and the names of years of the depression are mentioned above  the corresponding columns. The abscissa demarcates time relative to the critical time $t_{cr}$ (also indicated by a dashed gray line at zero), that is the time at which a depression is identified. The second dashed line indicates the time when the cyclone intensity began to weaken and dissipate.}
    \label{figsupp_NONdev_capeciTS}
\end{figure}

Here, we present the analysis for various cases when depressions formed cyclones (developing cases) in figure \ref{figsupp_dev_capeciTS}, and when depressions are formed that do not intensify to form cyclones (non-developing cases) in figure \ref{figsupp_NONdev_capeciTS}. The spatial average of network connectivity exhibits approximately diurnal oscillations in all cases (see row-I in figure \ref{figsupp_dev_capeciTS} and figure \ref{figsupp_NONdev_capeciTS}).

For developing cases, we find that the fractional intensity of CAPE $\langle CAPE \rangle _{HNS}/\langle CAPE\rangle$ increases much prior to the formation of the depression during most developing cases (see row-II in figure \ref{figsupp_dev_capeciTS}). On the other hand, we find that such increase in the fractional intensity of CAPE in regions of high node strengths is not evident for most non-developing cases(see row-II, in figure \ref{figsupp_NONdev_capeciTS}). Further, fractional intensity of convection inhibition $\langle CI \rangle _{HNS}/\langle CI\rangle$ increases in regions of high node strengths in developing cases (except cyclone Mocha), while there is no definite trend of CI related to regions of high-node strengths for non-developing cases (see row-III in figure \ref{figsupp_dev_capeciTS} and figure \ref{figsupp_NONdev_capeciTS}).  
Notice for cyclone Mocha, even though the increase in fractional intensity of CAPE is not as prominent, the fractional intensity of CI tends to zero (row-III, figure \ref{figsupp_dev_capeciTS}). Thus, in general, the net convection potential increases in regions associated with coherent vorticity dynamics.

Further the number density of locations associated with simultaneously high node strengths and CAPE ($f_{CAPE,NS}$) and CI ($f_{CI,NS}$) exhibit a synchronous co-evolution for developing cases (see row-IV in figure \ref{figsupp_dev_capeciTS}) which is not evident for non-developing cases (see row-IV in figure \ref{figsupp_NONdev_capeciTS}). The depression formed in 2022 March does exhibit a synchronous increase in $f_{CAPE,NS}$ and $f_{CI,NS}$, prior to the formation of the depression. However, such correlations are not sustained during the intensification of the depression (as also evident from the correlation statistics presented in the main text).

\section{Description of the supplementary animations of node strength evolution}\label{SIsec_describe_animatn}

We have provided animations to visualize the evolution of spatial distribution of node strengths across consecutive network time-frames for cyclone Amphan (2020) and cyclone Bangladesh (1991). Please see supplementary videos named \textit{`MOVIE-1.gif'} and \textit{`MOVIE-2.gif'}. The trajectory of the cyclone is demarcated by black dots and the location of the cyclone during a particular network time-frame ($29$ hours) is demarcated by red dots along the trajectory. The animations reveal local motion of patches of high node strengths over a few network time-stamps.

\begin{figure}[h!]
    \centering
    \includegraphics[width=12cm]{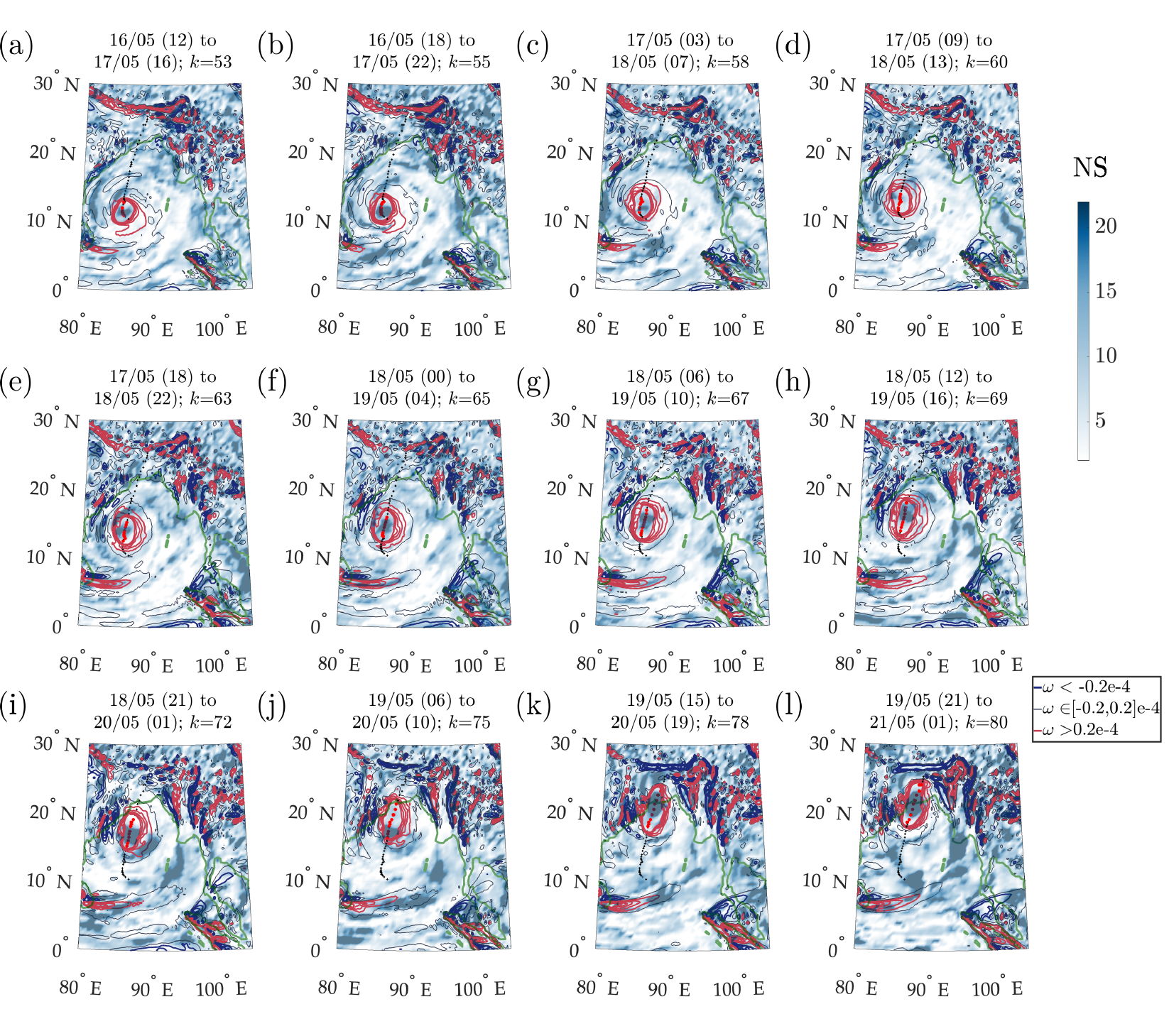}
    \caption{(a-l) Spatial distribution of node strengths $NS$ obtained from the network analysis \textbf{during the intensification and dissipation} of tropical cyclone Amphan (2020) during time-frame $k$, overlapped with contour lines representing isolines of relative vorticity ($\omega$) averaged over 29-hour period during the time-frame $k$. The time periods and network frame number ($k$) are mentioned in the title of each subplot. Strong positive and negative $\omega$ are represented by red and blue lines respectively, while black contours represent moderate values $\omega$. Patches of high node strengths appear to be arranged in a large-scale pattern that evolves with $k$. The trajectory of cyclone Amphan is represented by black dots beginning from the tropical depression state identified by IMD on May 16, 2020 at 00 hours. Larger red dots along the trajectory represent the location of cyclone Amphan during that time-frame.
  }
    \label{fig_NSvort_decay}
\end{figure}

Close to but prior to the development of a depression, we observe that these patches form extended patterns that appear in circular arrangements. The local motion of such patches then manifest as rotation of the large-scale pattern in cyclonic sense (anti-clockwise). Furthermore, as the network evolves, patches of high node strength form, appear to move and eventually dissipate. Even though some of the initially generated patches dissipate and new patches appear in adjacent location over several network time-frames, and the cyclonic motion of the large-scale pattern remains evident. The apparent motion is owing to percolation of local coherence of vorticity dynamics, and not due to material transport of fluid during the network time-frame, as explained in the main text. 

Patches appear along pockets in the wake of the cyclone, while extended structures are not evident during cyclone intensification. However, a circular arrangement of such patches of high node strength is still prominent during this period (evident in animation and figure \ref{fig_NSvort_decay}). As the cyclone weakens and dissipates, we observe that patches of high node strengths appear in extended arcs around the cyclone in its wake, and the circular arrangement of such patches disappears (figure \ref{fig_NSvort_decay}(i-l)).

\section{Supplementary Information}

\par{MOVIE 1:} Animation to visualize the evolution of spatial distribution of node strengths across consecutive network time-frames for cyclone Amphan (2020).

\par{MOVIE 2:} Animation to visualize the evolution of spatial distribution of node strengths across consecutive network time-frames for cyclone Bangladesh (1991).

\bibliography{refBMA}

\end{document}